\documentclass[10pt, aps, prc, superscriptaddress, amsmath, twocolumn,
 nofootinbib]{revtex4-1}

\usepackage{graphicx}
\usepackage[utf8]{inputenc}
\usepackage{hyperref}

\begin{document}

\title{Partial wave correlations with uncertainties in the Two Nucleon
 System}

\author{Adolfo Guevara Escalante}\email{adguevara@ugr.es}
\affiliation{Departamento de F\'{\i}sica At\'omica, Molecular y
  Nuclear and Instituto Carlos I de F{\'\i}sica Te\'orica y
  Computacional \\ Universidad de Granada, E-18071 Granada, Spain.}

\author{Rodrigo Navarro P\'erez}\email{rnavarroperez@sdsu.edu} \affiliation
 {Department of Physics. San Diego State University. 5500 Campanile Drive,
 San Diego, California 02182-1233, USA}

\author{Enrique Ruiz Arriola}\email{earriola@ugr.es} \affiliation
 {Departamento de F\'{\i}sica At\'omica, Molecular y Nuclear  and Instituto
 Carlos I de F{\'\i}sica Te\'orica y Computacional \\ Universidad de Granada,
 E-18071 Granada, Spain.}

\date{\today}

\begin{abstract} 
The NN scattering problem is usually analyzed in terms of partial waves and
the corresponding coupled channel phase-shifts and mixing angles, but the
available experiments induce correlations among the corresponding channels
with different quantum numbers. Based on the Granada-2013 database we analyze
the meaning and impact of those correlations taking into account both the
purely statistical ones reflecting the primary experimental data
uncertainties as well as the systematic ones exhibiting the ambiguities in
the form of the potential representing the unknown nuclear force for
distances below $3$fm. We find that the combined uncertainties, not only
display a dominance of systematic over statistical effects, but also show
that these correlations are almost compatible with zero.  These findings
support the frequent practice of determining potentials from separated
channel by channel direct fits to phase-shifts with the combined systematic
and statistical without the full fledged partial wave analysis (PWA) inferred
from experimental data but \emph{only} with much larger uncertainties.
\end{abstract}

\maketitle

\section{Introduction}

The determination of the NN scattering amplitude~\cite
{wolfenstein1952invariance} has traditionally been done in terms of a partial
wave expansion~\cite{ashkin1948neutron} and the corresponding phase-shifts
constrained by unitarity, particularly in the elastic energy regime located
below the pion production threshold. Early attempts used the phase-shifts
{\it directly} as fitting parameters at fixed energies for differential cross
sections and polarization observables~\cite{Stapp:1956mz} (see e.g. \cite
{phillips1959two, moravcsik1961theories} for reviews up to the late 50's).
Such a procedure proves crucial to unveil the most general NN potential~\cite
{okubo1958velocity} which is of concern to nuclear physics calculations, but
also induces correlations among the phase-shifts which need to be taken into
account for a faithful representation of the data within their given
{\it uncertainties}. The issue of correlations in this regard is rather old,
and within a phase-shift context it goes back to the mid 50's (see e.g.
~\cite{Anderson:1955zza} and \cite{MacGregor:1968zzd}). With some important
modifications, this is essentially the same procedure implemented over the
years by the countless number of attempts and still followed nowadays. A
historical overview up to 1989 has been reported~\cite{Machleidt:1989tm} and
the series of works~\cite{Stoks:1993tb, Stoks:1994wp, Wiringa:1994wb,
Machleidt:2000ge, Gross:2008ps, Perez:2013mwa, Perez:2013oba, Perez:2014yla,
Perez:2014waa, Perez:2016aol} describes statistically satisfactory
descriptions of the data since 1993 using phenomenological potentials. This
includes in particular some of the most up to date versions of the modern
chiral potentials~\cite{Epelbaum:2019kcf,RodriguezEntem:2020jgp}, which are
routinely used in {\it ab initio} calculations including 3-body chiral
interactions. (For recent and comprehensive reviews, see e.g.~\cite
{Machleidt:2017vls, RuizArriola:2019nnv, Perez:2020rtq} and references
therein). In this paper we undertake a detailed study of these statistical
correlations when from a large set of $8000$ pp and np scattering data below
$350$MeV LAB energy a subset of $6713$ of $3\sigma$-self consistent data
(the Granada 2013 database) is selected and fitted~\cite{Perez:2013mwa}, and
explore the multidimensional parameter space taking into account
{\it correlation uncertainties} which have been overlooked to the best of our
knowledge.

One common practice of nuclear theoreticians has been to propose NN potentials
and fitting to phase-shifts {\it separately}~\cite
{bryan1960new, Hamada:1962nq, Reid:1968sq, signell1969nuclear,
Lacombe:1975qr, Lagaris:1981mm, Wiringa:1984tg, Ordonez:1995rz,
NavarroPerez:2012qf, Epelbaum:2014sza} so that one could avoid undergoing a
full PWA. This is a rather convenient shortcut from a computational point of
view, and a good starting point to undertake a finer PWA, but such a scheme
is at odds with the existence of correlations. Actually, as it has been long
known~\cite{signell1963comparison, signell1964comparison} (see also
Ref.~\cite{Reid:1968sq} for an early discussion) and we have emphasized in
previous works~\cite{Perez:2014waa} a good fit to phase-shifts with an
acceptable confidence level does not imply a satisfactory description of the
complete scattering amplitude and casts doubts about the portability of the
partial wave analysis itself without explicitly quoting the correlations.
When such correlations are reported~\cite{MacGregor:1968zzd} correlated fits
become possible \cite{Nagels:1975fb, Nagels:1977ze, Nagels:1978sc}. However,
these correlations are subjected to uncertainties themselves. Therefore, by
undertaking the present correlation study we hope to give a specific answer
to the question under what conditions is the independent phase-shifts fit a
faithful description of the original scattering data including all sorts of
uncertainties.

Before embarking into the issue of correlations and their uncertainties, let
us review the nature of the problem in order to motivate our study. Much of
the current information about any theoretical analysis of NN scattering data
is often presented in terms of the corresponding coupled channel phase-shifts
and mixing angles which are determined by the conventional least squares
minimization method against the existing scattering data. The fitting method
acquires statistical meaning under certain conditions which can be checked
{\it a posteriori}~\cite{NavarroPerez:2014ihw, NavarroPerez:2014rvx}.
Typically these statistical methods allow for a determination and propagation
of the uncertainties and correlations of the fitting parameters and hence of
the corresponding phase-shifts.  The net result stressed in previous
works~\cite{Perez:2014waa, NavarroPerez:2016wqg, RuizSimo:2017anp} has been
that these statistical uncertainties turn out to be about an order of
magnitude smaller than the standard deviation between all the
(statistically equivalent, i.e. similar reduced $\chi^2$ values) PWA carried
out in the past. The reason may be attributed to the details on the
statistically equivalent interactions in the region below a relative
separation of 3 fm.  This is particularly significant; while phase-shifts are
not experimental observables themselves, they are regarded as
model-independent quantities which can be extracted from data. However, as we
will remind below, they turn out to be statistically-dependent objects as
inferred from data uncertainties and model-dependent due to the different
statistically equivalent fits based on different NN potentials~\footnote
{For NN potentials deduced from Quantum Field Theory at the hadronic level
this model dependence also covers finite cut-off regularization scheme
dependence or equivalently strong form factors due to short distance
singularities inherent in the perturbative evaluation of Feynman diagrams
involving a meson exchange picture (see e.g.  \cite{Machleidt:1989tm} for a
review). Modern effective field theory(EFT) approaches with suitable
counter-terms based on chiral symmetry(see e.g. \cite
{RodriguezEntem:2020jgp, Epelbaum:2019kcf} for recent reviews) also display a
scheme dependence which may be larger than nominally expected due to the need
of a finite regularization scheme~\cite{Carlsson:2015vda}.}. The existence of
correlations exhibits a level of redundancy and is quite natural within a
potential model approach to nuclear forces. It simply reflects the fact that
there are more phase-shifts than independent potential components according
to general symmetry principles as long as nucleons and pions are regarded as
elementary particles. Of course, one should not forget that on a more
fundamental level, the fact that QCD, the underlying theory of strong
interactions in terms of quarks and gluons for the two lightest $u,d$ flavors
essentially depends solely on two parameters which, in the isospin limit, can
be mapped at the hadronic level onto the pion weak decay $f_\pi$ and the pion
mass $m_\pi$ suggests that all correlations should be traced from the
underlying $f_\pi,m_\pi$ dependence. At the current experimental accuracy
those fundamental correlations are, however, practically invisible.  The same
is true at the fundamental level, despite encouraging progress on a
fundamental level within the lattice QCD approach to the NN problem~\cite
{Beane:2006mx, Aoki:2013tba} (see also e.g. \cite{Horz:2020zvv, Illa:2020nsi}
for recent studies and references therein and Ref.~\cite
{Drischler:2019xuo} for an overview). Thus we are left, for the time being,
with the phenomenological analysis.

The paper is structured as follows. In section~\ref{sec:ps} we review the
essential difference of the phase-shifts as primary or secondary parameters
and the corresponding fitting strategies as well as the implications for
statistical and systematic uncertainties. The necessary definitions on the
statistics of correlations are introduced in Section~\ref{sec:corr} and
particularized for our case. Our main numerical results are presented in
Section~\ref{sec:num} where we separate the study for both statistical and
systematic uncertainties at different confidence levels. Finally, in
Section~\ref{sec:conl} we come to the conclusions and provide an outlook.

\section{Phase-shifts as  primary or derived quantities}
\label{sec:ps}

While a procedure based in taking the phase-shifts as primary fitting
quantities facilitates enormously the direct evaluation of statistical
correlations through the corresponding covariance matrix at {\it fixed}
energy values, alternative procedures using NN potentials may be competitive
enough in terms of goodness of fit at {\it arbitrary} energy values but
complicate the uncertainty analysis of correlations. Thus, we find it
appropriate to ponder on the need of taking the phase-shifts as secondary
quantities, as it has become customary for the last fifty years
(see e.g. \cite{Arndt:1966in} and the discussion below). The reader familiar
with these issues may skip this section.

\subsection{Statement of the problem}

From a mathematical point of view the path from scattering data to the
scattering amplitude proves to be an unambiguous procedure~\cite
{Bystricky:1976jr} provided a complete set of measurements encompassing
differential cross sections and polarization observables at a particular
energy value are available~\cite{la1980scattering} (see \cite
{kamada2011determination} for an analytical solution). From the scattering
amplitudes the corresponding partial wave amplitudes and hence the
phase-shifts may be obtained.  While the situation of planning experiments
this way in a significant sample of measured energies would be the ideal one,
it has seldomly been applied in the region below pion production threshold of
most importance for theoretical nuclear physics in {\it ab initio}
calculations of binding energies of light nuclei. Instead rather fragmentary
intervals of energies, angles and measured observables carried out at
different laboratories are more frequently available to undertake large scale
analysis encompassing with as many compatible data as possible.

The formalism of NN PWA has been comprehensively reviewed in Refs.~\cite
{RuizArriola:2019nnv, Perez:2020rtq} with an emphasis put on the verification
vs falsification of statistical tests and the possibility of uncertainty
quantification and propagation. We refer to these papers for further details
and specific formulas in the general case corresponding to the scattering of
two spin 1/2 particles and the relation to experimental cross sections and
polarization observables as well as the modifications due to the tensor force
and the inclusion of determinant long range effects such as Coulomb, vacuum
polarization and relativistic corrections as well as charge dependent
One-Pion-Exchange potentials. For our purposes of illustrating the discussion
here, we may summarize the situation for the much simpler spinless spherical
local potential, $V(r)$, case where the only measurable observable would be
the differential cross section $\sigma (\theta,E) = |f(\theta,E)|^2$ with
$\theta$ the scattering angle and $E$ the scattering CM energy. The
scattering amplitude can be expanded in the conventional partial wave
expansion as
\begin{equation}
  f(\theta,E) = \sum_{l=0}^\infty (2l+1) \frac{e^{2i\delta_l (p)}-1}{2ip}P_l(\cos \theta). 
\end{equation}
Here, $P_l (z)$ are Legendre polynomials and $\delta_l(p)$ are the
phase-shifts depending on the CM momentum $p=\sqrt{2\mu E}$ with $\mu$ the
reduced mass. For the spherical potential case the total wave function can be
factorized as usual $\Psi(\vec x)= (u_l(r)/r) Y_{l,m} (\theta,\phi)$ with $Y_
{l,m} (\theta,\phi)$ the spherical harmonics and the phase shifts $\delta_l
(p)$ are computed by solving the reduced Schr\"odinger equation for the
reduced wave function $u_l(r)$ where
\begin{equation}
  -u_l''(r) + \left[ \frac{l(l+1)}{r^2} + 2\mu V(r) \right] u_l(r) = p^2 u_l (r), 
\end{equation}
with the asymptotic conditions (we assume  non-singular potentials $r^2  V
(r) \to 0$ as $r \to 0$)
\begin{equation}
  u_l(r) \underbrace{\to}_{r \to 0} r^{l+1}, \qquad 
  u_l(r) \underbrace{\to}_{r \to \infty} \sin \left( p r - \frac{l\pi}{2}+ \delta_l \right).
\end{equation}
For a potential with finite range $a$, the partial wave expansion is truncated
at about $l_{\rm max} \approx p_{\rm max} a$.

\subsection{Energy independent analysis}

The simplest situation corresponds to having complete data in a {\it single}
energy $E$ (or momentum $p$), namely $ \left(\sigma
(\theta_1, E) , \dots , \sigma(\theta_N, E) \right)$. In this case one can
determine the $l_{\rm max} \sim p a$ phase-shifts {\it directly} from the
data as fitting parameters $ (\delta_0 (E), \dots , \delta_{l_{\rm max}}
(E)) $ by minimizing
\begin{align}
  & \chi^2 ( \delta_1 (E), \dots , \delta_{l_{\rm max}} (E), Z) = \left(\frac{1-Z}{\Delta Z} \right)^2 + \nonumber \\
  & \sum_{i=1}^N \left[ \frac{ \sigma^{\rm exp} (\theta_i, E) - Z \sigma^{\rm th} (\theta_i,\delta_1 (E), \dots , \delta_{l_{\rm max}} (E))}{\Delta \sigma (\theta_i, E) } \right]^2.
\end{align} Here the
normalization $Z$ with estimated uncertainty $\Delta Z$ (provided by the
experimentalists) is {\it common} for {\it one} energy. Moreover, the matrix
error deduced as a second derivative with respect to the fitting parameters
allows to determine both the error and the correlations of different partial
waves at this same energy.  Thus, phase-shifts become ``experimental'' and
{\it model independent} observables, $ \delta_l^{\rm exp}
(E) \pm \Delta \delta_l^{\rm exp} (E) $ for $ l=0, \dots , l_{\rm max} $ at
this particular energy.

The energy independent analysis, despite being direct for extraction of
uncertainties and correlations suffers from some undesirable deficiencies.
Firstly, the number of active phase shifts increases with the energy, since
the maximal orbital angular momentum in the partial wave expansion is
typically $l_{\rm max} \sim 2 p_{\rm CM} /m_\pi$. Secondly, phase-shifts at
different energies may display trigonometric and unpleasant ambiguities,
although for nearby energies data may be extrapolated to a fixed energy.
Finally, every energy is treated independently, so that if we have less
experimental measurements than the number of necessary phase shifts, $N \le
l_{\rm max}$, a PWA becomes unfeasible. Although the energy independent
analysis is the only way to make a model independent PWA, it is no longer an
active strategy in part because of how the available experimental data have
been accumulated over the years.

\subsection{Energy dependent analysis}

If one has incomplete data for a fixed energy but a set of measurements at
several energies and angles $(\sigma(\theta_1,E_1) , \dots , \sigma
(\theta_N,E_N))$ one cannot generally determine phase-shifts $\delta_l
(E_i)$ at those energies because lack of data. Instead, a {\it model
dependent} interpolation with fitting parameters ${\bf p}$ in the energy is
needed, so that one has $\delta_l(E; {\bf p})$, i.e. the phase-shifts become
secondary or derived quantities~\footnote{A typical example is to take a
rational representation of the $K-$matrix, $p \cot \delta_l= \sum_{n=0}^K a_n
p^n / \sum_{n=0}^M b_n p^n$ where the corresponding coefficients build the
vector of fitting parameters ${\bf p}= (a_0, \dots, a_K; b_0, \dots, b_M)
$.}. Thus, one minimizes
\begin{align}
  \chi^2 ( {\bf p} , Z) &= \sum_{i=1}^N \left[ \frac{ \sigma (\theta_i,
    E_i)^{\rm exp} - Z \sigma^{\rm th} (\theta_i, E_i, {\bf p})}{
    \Delta \sigma (\theta_i, E_i) } \right]^2 \nonumber \\
    & + \left(\frac{1-Z}{\Delta Z} \right)^2 \, .
\end{align}
Different experiments have different normalization constants $Z$ so that
generally
\begin{equation}
  \chi^2 ( {\bf p} , Z_1, \dots Z_E ) = \sum_{i=1}^E \chi_i^2 ( {\bf p} , Z_i ). \label{eq:chinu}
\end{equation}
The outcome of this multi-energy analysis would be an error matrix for the
parameters ${\bf p}$ whence errors could be propagated to compute the error
matrix for the phase-shifts, using the standard covariance matrix approach.

The analysis pioneered by the Nijmegen group in the mid-nineties~\cite
{Stoks:1993tb} was the first example of a large scale PWA to NN data which a
statistically significant fit due to the inclusion of long range fine effects
and a meticulous selection of mutually consistent scattering data. This work
was partly followed by the Granada group and has allowed to pin down the NN
phase-shifts up to pion production threshold rather accurately~\cite
{Perez:2013jpa}. In the spinless case the Granada approach corresponds to a
separation of the potential into a well known quantum field theoretical piece
$V_{\rm QFT}(r)$ for the long range tail and a unknown short distance piece.
$V_{\rm QFT}(r)$ can be calculated by evaluating the corresponding Feynman
diagrams in perturbation theory. The short distance piece can be determined
by a PWA and is coarse grained on ``thick'' sampling points $r_n$ suitably
located by Dirac delta-shells as
\begin{equation}
  V_{\rm Short} (r)  =  \sum_n \Delta r V(r_n) \delta (r-r_n),
  \label{eq:vshort}  
\end{equation}
where $ r_n = n \Delta r $. Here $\Delta r \sim 1/p_{\rm max} \sim 0.6 $fm the
shortest de Broglie wavelength.
From the error matrix in the fitting parameters $V(r_i)$ one may propagate to
the error matrix of phase shifts.

In the general case with spin 1/2 particles, one has instead a set of
potential functions $V_i (r)$ associated to the general decomposition of the
potential in a given operator base $V (\vec x)=\sum_i O_i V_i (r)$
(see e.g. \cite{RuizArriola:2019nnv, Perez:2020rtq}) and as a consequence the
fitting parameters are given by $V_i(r_n)$. The upshot of the whole scheme is
the best fit of the Granada 2013 database with $N_{\rm Dat}=6741$ pp+np
selected scattering data below $350$MeV LAB energy with a total
$\chi^2=6855.50$ and $N_{\rm Par}=55$ fitting parameters, which corresponds
to a reduced $\chi^2$ per degrees of freedom of $\chi^2/
{\rm d.o.f}=1.025$~\cite{Perez:2016aol}. On the statistical level {\it most}
phases are determined by a 1 per mile or less accuracy, the main reason being
the strong constraints imposed by CD-One-Pion-Exchange interaction. For
recent reviews on uncertainties in the NN problem see e.g. \cite
{RuizArriola:2019nnv, Perez:2020rtq}.

While the model dependence due to the inclusion of suitable potentials is
generated by the need of intertwining fragmentary scattering measurements, it
has the further benefit that one can, in principle, use the fitted potentials
in {\it ab initio} nuclear structure calculations.  As we will see, it also
provides an additional source of uncertainty for those calculations, a
subject which has become in the last decade the corner stone of the
predictive power in Nuclear Physics.

\subsection{Systematic and statistical uncertainties}

The decomposition of the potential into an unknown short range part and a well
know long range part  complies with expected analytical properties of the
scattering amplitude in the absence of long distance electromagnetic
effects~\cite{RuizdeElvira:2018hsv} and provides a universal representation
for $V_{\rm QFT}(r)$. However, the representation of the short range part is
generally ambiguous, and the coarse grained representation, Eq.~(\ref
{eq:vshort}), while quite convenient and computationally cheap, is not unique
and several other functions have been proposed which are statistically
acceptable, i.e. they have $\chi^2/{\rm dof} \sim 1$. They are 7 pre-Granada
analyses starting with the Nijmegen bench-marking study~\cite
{Stoks:1993tb, Stoks:1994wp, Wiringa:1994wb, Machleidt:2000ge, Gross:2008ps}
the primary Granada 2013 $\chi^2$ analysis~\cite{Perez:2013mwa} and the
subsequent 5 Granada potentials~\cite{Perez:2013mwa, Perez:2013oba,
Perez:2014yla, Perez:2014waa}~\footnote{Our selection, based on comparison
with measurable data excludes several often used important potentials such as
the Extended soft core (see e.g. Ref.~\cite{Nagels:2014qqa}), might appear
to strict in cases where sometimes a qualitative agreement suffices .}.  The
differences in the phase-shifts are mainly attributed to a systematic
uncertainty in $V_{\rm Short}(r)$. In previous works~\cite
{Perez:2014waa, Perez:2016vzj, RuizSimo:2017anp, Perez:2020rtq,
RuizArriola:2019nnv} we have estimated systematic uncertainties taken a total
of $N=13$ analyses which have provided a satisfactory $\chi^2/{\rm d.o.f}$ at
the time of their fit. One important consequence of the Granada energy
dependent analysis corresponds to the out-coming uncertainty structure, where
it was found that the statistical uncertainties are about an order of
magnitude{\it smaller} than the systematic uncertainties~\cite
{Perez:2014waa, Perez:2016vzj, RuizSimo:2017anp, Perez:2020rtq,
RuizArriola:2019nnv}. This observation will be highly relevant for our
determination of systematic uncertainties of correlations.

So far, the dominance of systematic vs statistical uncertainties is a very
important but purely empirical observation based on combining the available
statistically equivalent PWA. We try to give here some {\it rationale} of the
situation. One may wonder how is it possible that statistically
equivalent model fits may yield different phase-shifts. In order to
understand this let us consider a situation where we could compare the energy
independent strategy, where phase-shifts are taken as primary fitting
parameters with the energy dependent one where the phase-shifts are obtained
from the minimizing potential parameters.

The corresponding phase-shifts will be denoted as $\delta_\alpha$ where
$\alpha=(E_n, l)$ runs over all the available energies and relevant angular
momenta. Thus, written in compact form, we have
\begin{equation}
  \min_{\delta} \chi^2 ( \delta ) \equiv \chi^2 (\delta^*),    
\end{equation}
where $\delta_\alpha^* $ are the minimizing phase-shifts whose statistical
uncertainties, $\Delta \delta_\alpha$ and correlation matrix can be obtained
from the standard covariance matrix.

In the energy dependent strategy we use a potential with parameters, which we
denote for short as $V_i$, we actually have that effectively the phase-shifts
become functions of the potential parameters $\delta_\alpha
(V) \equiv \delta_\alpha (V_1, \dots, V_ M)$ so that
\begin{equation}
  \chi^2 (V) = \chi^2 ( \delta (V)). 
\end{equation}
Thus, from a variational point of view, the minimization with respect to the
potential parameters operates in general as a {\it restriction}. Of course,
the introduction of arbitrary many parameters in the potential may produce
over-fitting which as we will see below is in fact not necessary. and hence
minimization means
\begin{equation}
  \min_{V} \chi^2 (\delta(V)) \equiv \chi^2 (\delta(V^*))  
  \ge \min_\delta \chi^2 (\delta) \equiv \chi^2 (\delta^*).
\end{equation}
Now, in the limit of small deviations we set 
\begin{equation}
  \delta_\alpha(V^*) =\delta_\alpha^* + \epsilon_\alpha, 
\end{equation}
and because of the stationary condition we get that the accuracy in the
minimum is quadratic in the  small deviation
\begin{equation}
  \chi^2 (\delta (V^*))- \chi^2 (\delta^*)=\chi^2 (\delta^* + \epsilon)- \chi^2 (\delta^*)=
      {\cal O}(\epsilon^2). 
\end{equation}
Thus, by construction we expect the energy dependent strategy to provide $
{\cal O}(\epsilon)$ phase-shifts for ${\cal O}(\epsilon^2)$ values of
$\chi^2$. In other words, using a potential we can get rather good fits with
not so precise phase-shifts. Any potential taken as a basis for the PWA will
induce a bias and we can only hope that by using different potentials with
different biases, the global bias will be removed in average.

In the above notation, the findings of our previous works correspond to the
fact that, if we take {\it different} potentials, say $V^{(1)}, V^{
(2)}, \dots $, in average $ \langle \epsilon_\alpha \rangle =0 $ but their
mean squared standard deviation ${\rm Std} (\epsilon_\alpha)$, which we call
the systematic uncertainty, is much larger than the statistical uncertainty,
$\Delta \delta_\alpha$, i.e.
\begin{equation}
  {\rm Std} (\epsilon_\alpha) \gg \Delta \delta_\alpha.
\end{equation}
The underlying reason of {\it why} there is such a big increase remains, to
our knowledge, a mystery. Nonetheless, as we will see later in this paper,
all these simple considerations have implications in the short-cut approach
of fitting nuclear potentials directly to phase-shifts extracted without
passing through the rather cumbersome approach of the PWA.

\section{Statistical definitions for correlations}
\label{sec:corr}

In order to assess numerically these correlations and their uncertainties we
rely on a few  definitions from statistics. We briefly introduce here the
main statistical quantities necessary to directly present our results.
However, we encourage the reader to review appendix \ref
{app:statistics_review} for a more in depth discussion of the theory on the
uncertainty estimate of the correlation coefficient. The appendix takes a
pedagogical approach since the theory, although a century old, does not
usually appear comprehensively in standard statistic textbooks (we take
Refs.
\cite{taylor1997introduction, evans2004probability, eadie2006statistical})

Our purpose in this work will be to estimate a confidence interval for the
population correlation $\rho$ among the  different NN partial waves at
different LAB energies. In practice we only have access to a finite sample of
pairs $(x_1, y_1) \dots(x_N, y_N)$ of size $N$ extracted from a bivariate
distribution $P(x,y)$. We define then standardized variables
\begin{equation}
  \hat x_i = \frac{x_i - \overline{x}}{s_x} \sqrt{\frac{N}{N-1}},
\end{equation}  
where the bar notation indicates the conventional sample mean and $s^2$
corresponds to the unbiased sample variance (see Eq.~(\ref{eq:sample_mean}) and
Eq.~(\ref{eq:sample_variance}) for specific definitions). The linear correlation
coefficient $r$, which is an estimator for $\rho$, is given by the sample
mean of the variable $ \xi_i = \hat x_i \hat y_i$
\begin{equation}
r \equiv  {\cal C}(\hat x, \hat y)= \overline{\xi}= \overline{ \hat x \hat y} =  \frac1{N} \sum_{i=1}^N \hat x_i \hat y_i
  \label{eq:corr}
\end{equation}
which fulfills the inequality $-1 \le {\cal C} \le 1$ with ${\cal C}=-1,0,1$
corresponding to full anti-correlation, full independence and full
correlation respectively of the sets $\{\hat x_1 , \dots, \hat x_N \}$ and
$ \{\hat y_1 , \dots, \hat y_N \}$. Of course, the correlation coefficient
depends on the sample of size $N$ and hence different extractions will
produce different values of $r$ so that it will eventually provide a
distribution $P_{\rho, N}(r)$ due to the finite sample size.

\begin{figure}
 \begin{center}
 \includegraphics[width=\linewidth]{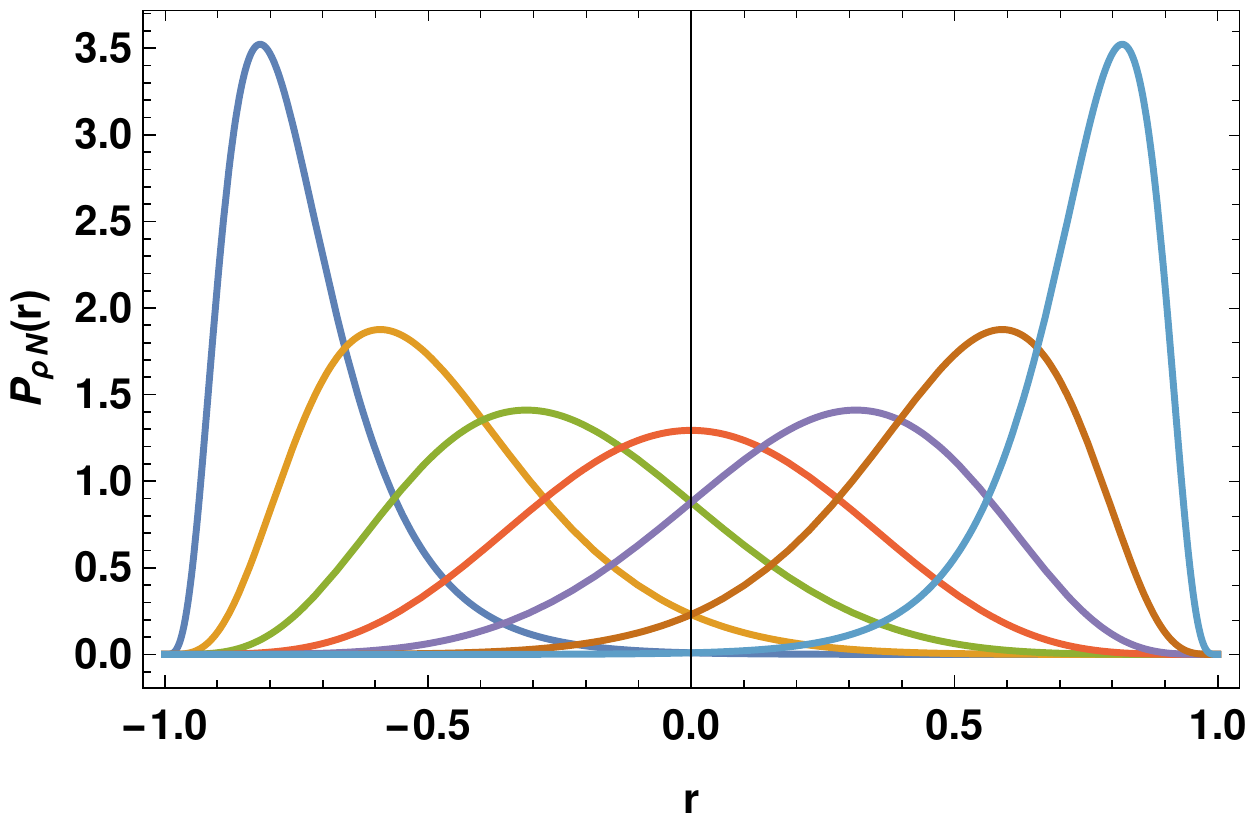}
  \end{center}
 \caption{Theoretical distribution function of the correlation
   coefficient $r$ for a sample of size $N=13$ generated from a
   bivariate Gaussian population with theoretical correlation
   coefficients (from left to right)
   $\rho=-0.75,-0.5,-0.25,0,0.25,0.50,0.75$.}\label{Fig:f(r)}
\end{figure}

For the particular case of a bivariate Gaussian distribution the resulting
$P_{\rho, N}(r)$ has been derived long ago and can be found in Eq.~(\ref
{eq:correlation_dist}). Fig.~\ref{Fig:f(r)} shows the corresponding
distribution $P_{\rho,N}(r)$ for a few values of the correlation coefficient
and in the particular case $N=13$ which will be our main interest here.  As
we can clearly see, rather large empirical values , $|r|\ge 0.5 $ would be
needed if one claims that $\rho \neq 0$ significantly. 

\begin{figure}
\begin{center}
 \includegraphics[width=\linewidth]{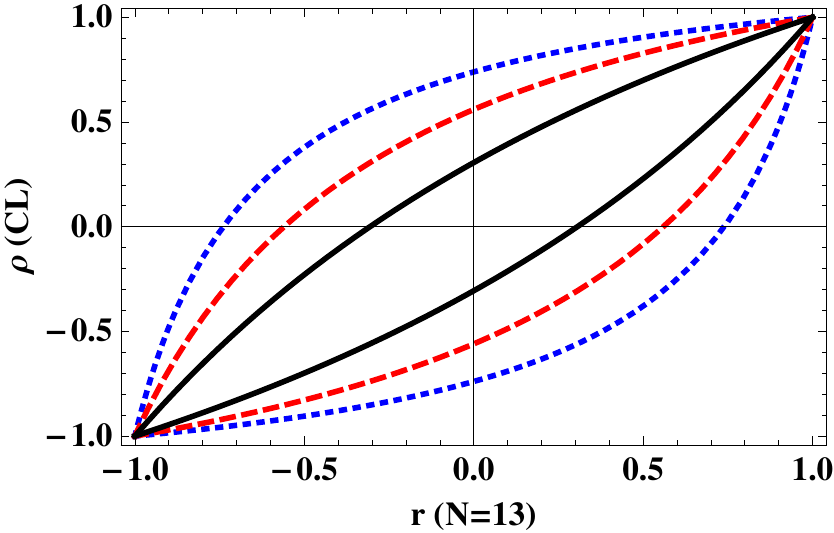}
  \end{center}
  \caption{Estimated confidence limits for the correlation $\rho$
   as a function of the empirical correlation coefficient of a sample
   of size $N=13$ for 1,2 and 3 standard deviations corresponding to
   $68\%,95\%,99\%$ confidence level respectively. }\label{Fig:rho(r)}
\end{figure}

According to a naive application of the central limit theorem, for $N \gg 1$,
the distribution becomes a Gaussian with mean $\rho$ and standard deviation
$\sqrt{(1+\rho^2)/N}$. In that case the formula involving the sample variance
can be used directly
 \begin{equation}
   \rho = \overline{\hat x \hat y } \pm \frac1{\sqrt{N}} \left[
     \overline{  ( \hat x \hat y -  \overline{\hat x \hat y})^2 
     }
     \right]
\label{eq:r-cl}   
\end{equation}
which for our case $N=13$ works rather well for $|\rho| \ll 1$, but fails for
sizeable correlations due to the asymmetric shape of the distribution. Eq.
(\ref{eq:r-cl}) can erroneously give estimates of $r$ which fall outside the
interval $[-1,1]$. A way to address the general asymmetric situation is by
recurring to the central limit theorem, in terms of the Fisher transform,
which for large $N$ behaves as Gaussian variable
\begin{equation}
  z= \frac12 \ln \frac{1+r}{1-r} =   \frac12 \ln \frac{1+\rho}{1-\rho} + 
\frac{\xi}{\sqrt{N-3}}  
\end{equation}
with $\xi \in N[0,1]$. Thus $z $ has a mean $\mu_z= \tanh^{-1} \rho$ and
variance $\sigma_z=1/\sqrt{N-3}$. For our $N=13$ case this formula works well
enough (it can hardly be distinguished in Fig.~\ref{Fig:f(r)} so that we do
not plot it) for the asymmetric case and higher order corrections may be
found in Ref.~\cite{vrbik2005population}. Therefore we have the ranges
\begin{equation}
  \rho = \tanh \left\{ \tanh^{-1}(r) \pm  \frac{z}{\sqrt{N-3}} \right\}
\label{eq:r-CL}  
\end{equation}
where $z=1,2,3$ corresponds to $68\%,95\%,99\%$ confidence level respectively.
These three CL bands are displayed for illustration purposes in Fig.~\ref
{Fig:rho(r)}.
       
\section{NN Partial waves correlations}
\label{sec:num}

We may now proceed directly to evaluate the corresponding linear correlation
coefficients. For any pair of partial waves, say $\delta_\alpha$ and
$\delta_\beta$ and for a given LAB energy we proceed by computing the sample
correlation ${\cal C} (\hat \delta_\alpha , \hat \delta_\beta)$ from Eq.~\ref
{eq:corr} and assign the error according to Eq.~\ref{eq:r-CL} for $z=1,2$
corresponding to $68\%$ and $95\%$ confidence level. As already mentioned we
will address separately both the correlations due to purely statistical
origin, i.e. directly stemming from the experimental measurements, and our
estimates of systematic origin corresponding to the different representations
and parameterizations of the interaction and reflecting inherent ambiguities
in the scattering problem.

In our study of the 13 PWA fits we will segregate the 6 Granada potentials
from the remaining 7 previous phenomenological approaches. While all the
potentials share the necessary long range effects to provide statistically
significant fits at their time, this separation makes sense for a variety of
reasons. Firstly, all the Granada analyses are explicitly provided in
published form with statistical error bars, a practice which only applies to
the original and bench-marking Nijmegen analysis (but unfortunately not to
their subsequent high-quality potentials).  Secondly, the selected
database is the same. A third reason is that the strict separation between
$V_{\rm QFT}(r)$ and $V_{\rm Short}(r)$ is applied in all 6 Granada cases.
Finally, it is encouraging that grouping the Granada potentials separately,
there seems to be no particular bias as compared to the precedent and quite
diverse PWA using the original and 40$\%$ smaller 1993 Nijmegen database.

\begin{figure*}
 \centering\includegraphics[width=.95\linewidth]{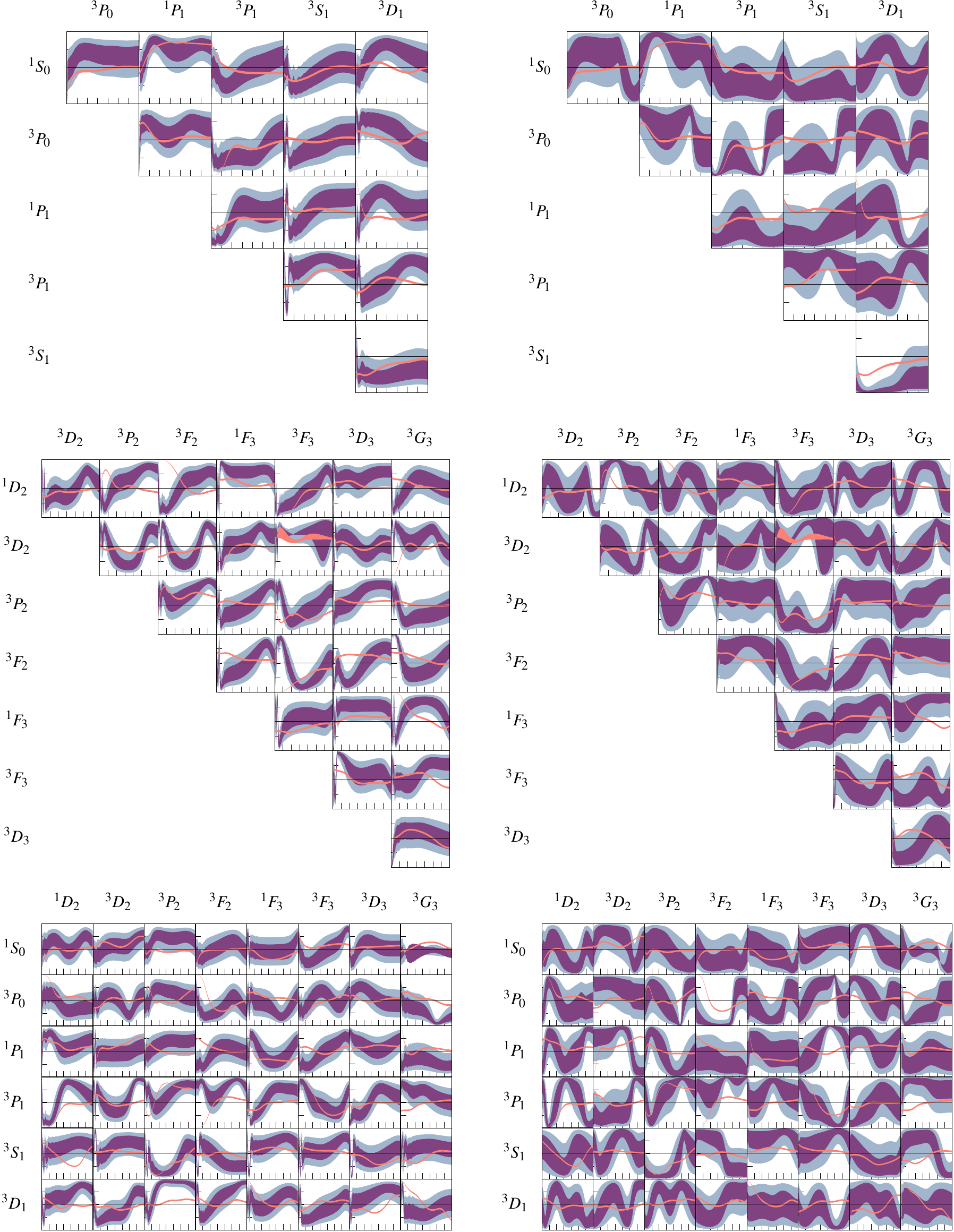}
  \caption{(Color online) Partial wave correlation coefficients $-1
    \le r \le 1 $ with confidence level as a function of the nucleon LAB
     energies for $0 \le T_{\rm LAB} \le 350$MeV. The larger (light blue)
     band corresponds to the systematic uncertainty at the $2\sigma$
     confidence level. The medium band (purple) corresponds to the systematic
     uncertainty at the $1\sigma$ confidence level. The smaller (orange) band
     corresponds to the statistical uncertainty at the $1\sigma$ confidence
     level as described in the text. Top: Central phases with total angular
     momentum $J \le 1$. Middle: Peripheral phases with $J\ge 2$. Bottom:
     Central-peripheral correlations. We show the results for the 13 HQ
     potentials quoted in the text (left) and the 6 Granada potentials
     (right).}\label{fig:correlat}
\end{figure*}

We consider correlations among all partial waves with total angular momentum
$J \le 3$, corresponding to waves from $^1S_0$ to $^3G_3$. Our results for
the linear correlation coefficients among different partial waves are shown
in Fig.~\ref{fig:correlat} where we show the $1\sigma$ and $2\sigma$
confidence levels according to the Fisher transformation mapping specified in
Eq.~(\ref{eq:r-CL}).  For a better comparison we display them in a block form
resembling the (symmetric) correlation matrix and splitting the upper
diagonal involving the central phases, which have a total angular momentum
$J \le 1$ and the peripheral phases $J \ge 2$. In this representation the
off-diagonal block corresponds to the correlation between central and
peripheral waves.  The thin band corresponds to the purely statistical
correlations of the Granada 2013 partial wave analysis whereas the thicker
bands represent either the 13 high-quality potentials developed since the
Nijmegen analysis (left panels) as well as the 6 Granada interactions
(right panels). In all cases we take as the x-axis the nucleon LAB energies
for $0 \le T_{\rm LAB} \le 350$MeV (ticks represent each multiples of 50 MeV)
and the y-axis as the correlation coefficient in the range $-1 \le r \le 1$.

\subsection{Statistical correlations}


Correlations are generated using the bootstrap method, where the 6713 np+pp
experimental results corresponding to 6173 scattering data containing
differential cross sections and polarization asymmetries are replicated
$N=1000$ times following Ref.~\cite{Perez:2014jsa} by performing a Gaussian
fluctuation on the existing data and taking the CD-OPE with delta-shells and
the pion-nucleon-nucleon coupling constants as fitting parameters and
minimizing the $\chi^2$ for each the $N=1000$ replicas. This approach
produces a non-parametric multidimensional distribution of parameters which
is not necessarily Gaussian (see~\cite{Perez:2014jsa} for illustrations) but
also a $N=1000$ sample of population of phase-shifts at any single LAB energy
value whence their mutual correlations and the corresponding uncertainties
using the formulas in the previous section can be obtained.

For a finite range potential with a sharp boundary, the partial wave expansion
has a cut-off in the maximal angular momentum $J_{\rm max}$ which roughly and
in a semi-classical picture corresponds to the maximal impact parameter
consistent with the occurrence of a collision, namely $b \lesssim r_c$. The
exponential fall-off of the OPE potential above a certain cut-off radius
implies that larger angular momenta than this $J_{\rm max}$ are mainly
determined by the OPE potential tail.

According to the Granada analysis, the partial waves may roughly be divided
into active and passive channels corresponding to the low partial waves
actually involving the fitting parameters and the higher partial waves
mainly, but not fully, determined by the CD-OPE potential. We naturally
expect the peripheral partial waves to be strongly correlated because they
behave approximately in a perturbative manner and are determined by a unique
Yukawa-like function.

\subsection{Systematic correlations}

Our way of handling systematic correlations is to regard the very choice of
the short range potential as random and treat them as a sample of a
population of possible potential choices. This regards a total of $N=13$ PWA
carried out in the last 25 years and which have been successful within a
statistical point of view, namely the pioneering Nijmegen analysis and a
total of 6 Granada potentials.  One common feature of all the fitting
potentials is that they contain exactly the same CD-OPE potential starting at
distances larger than $3$fm and all other EM corrections such as
relativistic, magnetic moments and vacuum polarization.

The results are depicted in Fig.~\ref{fig:correlat} and, consistent with
previous studies, systematic uncertainties are much larger than statistical
uncertainties. Most remarkable is the fact that many of the apparently
significant statistical correlations turn into lack of correlations within
uncertainties. Thus, if we accept the spread of phase-shifts the $N=13$
potentials as a lower bound on the current uncertainty, one may take most of
partial waves as {\it uncorrelated} and one may proceed to fit different
partial wave channels independently of each other.

\section{Conclusions}
\label{sec:conl}

In this paper we have analyzed the correlations and their uncertainties among
the different partial waves in NN scattering below 350 MeV stemming either
from statistical data uncertainties or a lack of knowledge on the interaction
at short distances.

One of the direct applications of the results found here is that the naive
approach of fitting NN phase-shifts to fix nuclear forces in potential models
without undertaking a large scale partial wave analysis to experimental data
may, to some extent, be justified.  The reason is that regarding the largest
source of uncertainty which corresponds to our current deficient
representation of the NN interaction below 3fm phase-shifts in different
partial waves are {\it un-correlated} in the entire energy range and within
the corresponding uncertainties. The price to pay, however, is that this
implies dealing an order of magnitude larger uncertainties in the phase
shifts.

Ignoring correlations in the fits is then justified only at the $2\sigma$
level, namely when we {\it enlarge} their mean standard deviation of the 13
potentials, which roughly corresponds to 20 times larger uncertainties than
in the PWA fitting and selecting the $3\sigma$-self consistent 2013 Granada
database. In practice this adds further uncertainties, to the already
existing ones, to {\it ab initio} nuclear structure calculations in terms of
potentials obtained by this procedure.

\begin{acknowledgments}
This work is supported by the Spanish MINECO and European FEDER funds
(grant FIS2017-85053-C2-1-P) and Junta de Andalucía (grant FQM-225)
\end{acknowledgments}

\appendix

\section{Statistics of correlations}

\label{app:statistics_review}

We start by considering a uni-variate and normalized distribution $P(x)$. For
such a distribution we have the expectation value of a function $O
(x)$ defined as
\begin{equation}
  \langle O \rangle = \int_{-\infty}^\infty dx O(x) P(x) 
\end{equation}
where $\langle 1 \rangle=1$ is the normalization condition, and $\mu_x
\equiv \langle x \rangle $ is the population mean and $\sigma_x \equiv
\langle (x- \langle x \rangle)^2 \rangle = \langle x^2 \rangle -
\langle x \rangle^2$ is the population mean squared distribution. For a sample
$(x_1, \dots, x_N)$ of size $N$ extracted from this distribution $P$ we
introduce the conventional statistical mean and variance as unbiased
estimators
\begin{align}
\overline{x} &= \frac{1}{N} \sum_{i=1}^N x_i \\ 
s_x^2 &=\frac{N}{N-1} \overline{ (x- \overline{x})^2 } 
\end{align}
which fulfill the properties 
\begin{align}
  \label{eq:sample_mean}
  \langle   \overline{x} \rangle &= \langle x \rangle = \mu_x\\
  \label{eq:sample_variance}
  \langle   s_x^2 \rangle &= \langle (x- \langle x \rangle)^2 \rangle = \sigma_x^2
\end{align}
The sample mean $\bar x$ is itself a random variable since any different
extraction of size $N$ will generally produce a different result. The
corresponding distribution function fulfills for large samples, $N \gg 1$,
the well-known central limit theorem
\begin{equation}
  \label{eq:sample_distribution}
  P_N(z) = \prod_{i=1}^N \int dx_i P(x_i) \delta (z-\bar x) \to
  \frac{e^{-\frac12 \frac{(z-\mu)^2}{\sigma^2/N}}}{\sqrt{2\pi \sigma^2/N}}
\end{equation}
where $\delta(x)$ is the usual Dirac delta function. This result is customarily
summarized by stating that $\bar x = \mu \pm \sigma/\sqrt{N}$ with a $68\%$
confidence level. Inverting this relation one finds an estimate of the
population mean in terms of the sample mean and sample variance, $\mu = \bar
x \pm s_x /\sqrt{N}$ 

The corresponding extension to a normalized joint bivariate
probability function $P(x, y )$ is straightforward and we define accordingly
the expectation value as
\begin{equation}
  \langle O \rangle  = \int dx d y  \,  O(x, y) P(x,y)  
\end{equation}
For our discussion it will be useful to define standardized variables
\begin{equation}
  \hat x = \frac{x - \mu_x}{\sigma_x} \, , \quad
    \hat y = \frac{y - \mu_y}{\sigma_y}
\end{equation}
Thus, by construction we have 
\begin{align}
  \langle \hat x \rangle &= \langle \hat y \rangle =0 \\ \langle \hat
  x^2 \rangle &= \langle \hat y^2 \rangle =1
\end{align}
and introduce the linear correlation coefficient $\rho$ and its variance
$\sigma_\rho^2$
\begin{align}
  \rho &= \langle \hat x
  \hat y\rangle = \frac{\sigma_{xy}}{\sigma_x \sigma_y}\\ \sigma_\rho^2 &= \langle ( \hat x \hat y - \langle
  \hat x \hat y \rangle)^2 \rangle
\end{align}
where $\sigma_{xy}= \langle( x-\mu_x) (y - \mu_y) \rangle$ is the covariance
of $x$ and $y$. 

The correlation $\rho$ between the $x$ and $y$ variables can be estimated
taking a finite sample of size $N$ and calculating the correlation
coefficient defined in Eq.~(\ref{eq:corr}) of the main text. Similarly to the
sample mean $\bar x$, $r$ is also a random variable since a different sample
of the same size $N$ will result, generally, in a different value for $r$.
Therefore, the distribution function for the linear correlation coefficient
of $N$ pairs is given by the $\delta-$constrained integral
\begin{equation}
  P_N(r) =  \left\{ \prod_{i=1}^N \int d\hat x_i d\hat y_i P(\hat x_i , \hat y_i)  \right\} 
  \delta(r - {\cal C}(\hat x,\hat y))   
\end{equation}
which fulfills the proper normalization condition
\begin{equation}
\int_{-1}^1 dr   P_N(r)  = 1 \, . 
\end{equation}
For a standardized bivariate Gaussian distribution of the form 
\begin{equation}
P(x,y)= \frac1{2\pi \sqrt{1-\rho^2}} e^{-\frac{x^2 + y^2 -2 \rho xy }{2(1-\rho^2)}}
\end{equation}
the distribution function $P_N(r)$ has been evaluated analytically long
ago~\cite{hotelling1953new}
\begin{align}
  \label{eq:correlation_dist}
  P_{\rho,N}(r) &= \frac{(N-1)}{\sqrt{2 \pi }} \frac{\Gamma (N)}{\Gamma \left(N+\frac{1}{2}\right)}  \left(1-r^2\right)^{\frac{N-3}{2}} \left(1-\rho ^2\right)^{N/2} \nonumber \\
  &\times   (1 - r \rho )^{\frac{1}{2}-n} F \left(  \frac12;\frac12;N+\frac{1}{2};\frac{1}{2} (r \rho +1)\right)
\end{align}
where $\Gamma(x)$ is Euler's gamma function and $F(a,b,c,x)$ is the
hypergeometric function, whose power series around $x=0$ reads,
\begin{equation}
  F(a,b,c,x)= \sum_{n=0}^\infty \frac{(a)_n (b)_n}{(c)_n} \frac{x^n}{n!}
\end{equation}
and $(a)_n= a(a+1) \cdots (a+n-1)$ is the Pochhammer symbol. For example, for
uncorrelated Gaussian distributions , corresponding to $\rho=0$ , the exact result
simplifies to
\begin{equation}
P_{0,N}(r)= \frac{\left(1-r^2\right)^{\frac{N-3}{2}} \Gamma \left(\frac{N}{2}\right)}{\sqrt{\pi } \Gamma
   \left(\frac{N-1}{2}\right)}
\end{equation}
In the general $\rho \neq 0$ case one has
\begin{equation}
  \langle r \rangle= \rho \, , \quad  \sigma_r^2= \langle r^2 \rangle - \langle r \rangle^2= \frac{1+\rho^2}{N}
\end{equation}
which means that for a finite sample we may observe finite correlations
$r \in (-1,1)/\sqrt{n}$ even though the original population is free of them
with $\sim 68\%$ confidence level.  For example, for $N=16$ a correlation
coefficient of $|r| \le 0.25$ is largely compatible with no correlation.
Plots of the $P_{\rho,N=13}(r)$ distribution for different values of $\rho$
can be seen in Fig.~\ref{Fig:f(r)} of the main text.

\bibliography{revised-NN_PW_correlations.bib}

\begin{thebibliography}{63}%
\makeatletter
\providecommand \@ifxundefined [1]{%
 \@ifx{#1\undefined}
}%
\providecommand \@ifnum [1]{%
 \ifnum #1\expandafter \@firstoftwo
 \else \expandafter \@secondoftwo
 \fi
}%
\providecommand \@ifx [1]{%
 \ifx #1\expandafter \@firstoftwo
 \else \expandafter \@secondoftwo
 \fi
}%
\providecommand \natexlab [1]{#1}%
\providecommand \enquote  [1]{``#1''}%
\providecommand \bibnamefont  [1]{#1}%
\providecommand \bibfnamefont [1]{#1}%
\providecommand \citenamefont [1]{#1}%
\providecommand \href@noop [0]{\@secondoftwo}%
\providecommand \href [0]{\begingroup \@sanitize@url \@href}%
\providecommand \@href[1]{\@@startlink{#1}\@@href}%
\providecommand \@@href[1]{\endgroup#1\@@endlink}%
\providecommand \@sanitize@url [0]{\catcode `\\12\catcode `\$12\catcode
  `\&12\catcode `\#12\catcode `\^12\catcode `\_12\catcode `\%12\relax}%
\providecommand \@@startlink[1]{}%
\providecommand \@@endlink[0]{}%
\providecommand \url  [0]{\begingroup\@sanitize@url \@url }%
\providecommand \@url [1]{\endgroup\@href {#1}{\urlprefix }}%
\providecommand \urlprefix  [0]{URL }%
\providecommand \Eprint [0]{\href }%
\providecommand \doibase [0]{http://dx.doi.org/}%
\providecommand \selectlanguage [0]{\@gobble}%
\providecommand \bibinfo  [0]{\@secondoftwo}%
\providecommand \bibfield  [0]{\@secondoftwo}%
\providecommand \translation [1]{[#1]}%
\providecommand \BibitemOpen [0]{}%
\providecommand \bibitemStop [0]{}%
\providecommand \bibitemNoStop [0]{.\EOS\space}%
\providecommand \EOS [0]{\spacefactor3000\relax}%
\providecommand \BibitemShut  [1]{\csname bibitem#1\endcsname}%
\let\auto@bib@innerbib\@empty
\bibitem [{\citenamefont {Wolfenstein}\ and\ \citenamefont
  {Ashkin}(1952)}]{wolfenstein1952invariance}%
  \BibitemOpen
  \bibfield  {author} {\bibinfo {author} {\bibfnamefont {L.}~\bibnamefont
  {Wolfenstein}}\ and\ \bibinfo {author} {\bibfnamefont {J.}~\bibnamefont
  {Ashkin}},\ }\href@noop {} {\bibfield  {journal} {\bibinfo  {journal}
  {Physical Review}\ }\textbf {\bibinfo {volume} {85}},\ \bibinfo {pages} {947}
  (\bibinfo {year} {1952})}\BibitemShut {NoStop}%
\bibitem [{\citenamefont {Ashkin}\ and\ \citenamefont
  {Wu}(1948)}]{ashkin1948neutron}%
  \BibitemOpen
  \bibfield  {author} {\bibinfo {author} {\bibfnamefont {J.}~\bibnamefont
  {Ashkin}}\ and\ \bibinfo {author} {\bibfnamefont {T.-Y.}\ \bibnamefont
  {Wu}},\ }\href@noop {} {\bibfield  {journal} {\bibinfo  {journal} {Physical
  Review}\ }\textbf {\bibinfo {volume} {73}},\ \bibinfo {pages} {973} (\bibinfo
  {year} {1948})}\BibitemShut {NoStop}%
\bibitem [{\citenamefont {Stapp}\ \emph {et~al.}(1957)\citenamefont {Stapp},
  \citenamefont {Ypsilantis},\ and\ \citenamefont {Metropolis}}]{Stapp:1956mz}%
  \BibitemOpen
  \bibfield  {author} {\bibinfo {author} {\bibfnamefont {H.}~\bibnamefont
  {Stapp}}, \bibinfo {author} {\bibfnamefont {T.}~\bibnamefont {Ypsilantis}}, \
  and\ \bibinfo {author} {\bibfnamefont {N.}~\bibnamefont {Metropolis}},\
  }\href {\doibase 10.1103/PhysRev.105.302} {\bibfield  {journal} {\bibinfo
  {journal} {Phys. Rev.}\ }\textbf {\bibinfo {volume} {105}},\ \bibinfo {pages}
  {302} (\bibinfo {year} {1957})}\BibitemShut {NoStop}%
\bibitem [{\citenamefont {Phillips}(1959)}]{phillips1959two}%
  \BibitemOpen
  \bibfield  {author} {\bibinfo {author} {\bibfnamefont {R.}~\bibnamefont
  {Phillips}},\ }\href@noop {} {\bibfield  {journal} {\bibinfo  {journal}
  {Reports on Progress in Physics}\ }\textbf {\bibinfo {volume} {22}},\
  \bibinfo {pages} {562} (\bibinfo {year} {1959})}\BibitemShut {NoStop}%
\bibitem [{\citenamefont {Moravcsik}\ and\ \citenamefont
  {Noyes}(1961)}]{moravcsik1961theories}%
  \BibitemOpen
  \bibfield  {author} {\bibinfo {author} {\bibfnamefont {M.~J.}\ \bibnamefont
  {Moravcsik}}\ and\ \bibinfo {author} {\bibfnamefont {H.~P.}\ \bibnamefont
  {Noyes}},\ }\href@noop {} {\bibfield  {journal} {\bibinfo  {journal} {Annual
  review of nuclear science}\ }\textbf {\bibinfo {volume} {11}},\ \bibinfo
  {pages} {95} (\bibinfo {year} {1961})}\BibitemShut {NoStop}%
\bibitem [{\citenamefont {Okubo}\ and\ \citenamefont
  {Marshak}(1958)}]{okubo1958velocity}%
  \BibitemOpen
  \bibfield  {author} {\bibinfo {author} {\bibfnamefont {S.}~\bibnamefont
  {Okubo}}\ and\ \bibinfo {author} {\bibfnamefont {R.~E.}\ \bibnamefont
  {Marshak}},\ }\href@noop {} {\bibfield  {journal} {\bibinfo  {journal}
  {Annals of Physics}\ }\textbf {\bibinfo {volume} {4}},\ \bibinfo {pages}
  {166} (\bibinfo {year} {1958})}\BibitemShut {NoStop}%
\bibitem [{\citenamefont {Anderson}\ \emph {et~al.}(1955)\citenamefont
  {Anderson}, \citenamefont {Davidon}, \citenamefont {Glicksman},\ and\
  \citenamefont {Kruse}}]{Anderson:1955zza}%
  \BibitemOpen
  \bibfield  {author} {\bibinfo {author} {\bibfnamefont {H.}~\bibnamefont
  {Anderson}}, \bibinfo {author} {\bibfnamefont {W.}~\bibnamefont {Davidon}},
  \bibinfo {author} {\bibfnamefont {M.}~\bibnamefont {Glicksman}}, \ and\
  \bibinfo {author} {\bibfnamefont {U.}~\bibnamefont {Kruse}},\ }\href
  {\doibase 10.1103/PhysRev.100.279} {\bibfield  {journal} {\bibinfo  {journal}
  {Phys. Rev.}\ }\textbf {\bibinfo {volume} {100}},\ \bibinfo {pages} {279}
  (\bibinfo {year} {1955})}\BibitemShut {NoStop}%
\bibitem [{\citenamefont {MacGregor}\ \emph {et~al.}(1968)\citenamefont
  {MacGregor}, \citenamefont {Arndt},\ and\ \citenamefont
  {Wright}}]{MacGregor:1968zzd}%
  \BibitemOpen
  \bibfield  {author} {\bibinfo {author} {\bibfnamefont {M.~H.}\ \bibnamefont
  {MacGregor}}, \bibinfo {author} {\bibfnamefont {R.~A.}\ \bibnamefont
  {Arndt}}, \ and\ \bibinfo {author} {\bibfnamefont {R.~M.}\ \bibnamefont
  {Wright}},\ }\href {\doibase 10.1103/PhysRev.169.1128} {\bibfield  {journal}
  {\bibinfo  {journal} {Phys. Rev.}\ }\textbf {\bibinfo {volume} {169}},\
  \bibinfo {pages} {1128} (\bibinfo {year} {1968})}\BibitemShut {NoStop}%
\bibitem [{\citenamefont {Machleidt}(1989)}]{Machleidt:1989tm}%
  \BibitemOpen
  \bibfield  {author} {\bibinfo {author} {\bibfnamefont {R.}~\bibnamefont
  {Machleidt}},\ }\href@noop {} {\bibfield  {journal} {\bibinfo  {journal}
  {Adv. Nucl. Phys.}\ }\textbf {\bibinfo {volume} {19}},\ \bibinfo {pages}
  {189} (\bibinfo {year} {1989})}\BibitemShut {NoStop}%
\bibitem [{\citenamefont {Stoks}\ \emph {et~al.}(1993)\citenamefont {Stoks},
  \citenamefont {Klomp}, \citenamefont {Rentmeester},\ and\ \citenamefont
  {de~Swart}}]{Stoks:1993tb}%
  \BibitemOpen
  \bibfield  {author} {\bibinfo {author} {\bibfnamefont {V.}~\bibnamefont
  {Stoks}}, \bibinfo {author} {\bibfnamefont {R.}~\bibnamefont {Klomp}},
  \bibinfo {author} {\bibfnamefont {M.}~\bibnamefont {Rentmeester}}, \ and\
  \bibinfo {author} {\bibfnamefont {J.}~\bibnamefont {de~Swart}},\ }\href
  {\doibase 10.1103/PhysRevC.48.792} {\bibfield  {journal} {\bibinfo  {journal}
  {Phys. Rev. C}\ }\textbf {\bibinfo {volume} {48}},\ \bibinfo {pages} {792}
  (\bibinfo {year} {1993})}\BibitemShut {NoStop}%
\bibitem [{\citenamefont {Stoks}\ \emph {et~al.}(1994)\citenamefont {Stoks},
  \citenamefont {Klomp}, \citenamefont {Terheggen},\ and\ \citenamefont
  {de~Swart}}]{Stoks:1994wp}%
  \BibitemOpen
  \bibfield  {author} {\bibinfo {author} {\bibfnamefont {V.}~\bibnamefont
  {Stoks}}, \bibinfo {author} {\bibfnamefont {R.}~\bibnamefont {Klomp}},
  \bibinfo {author} {\bibfnamefont {C.}~\bibnamefont {Terheggen}}, \ and\
  \bibinfo {author} {\bibfnamefont {J.}~\bibnamefont {de~Swart}},\ }\href
  {\doibase 10.1103/PhysRevC.49.2950} {\bibfield  {journal} {\bibinfo
  {journal} {Phys. Rev. C}\ }\textbf {\bibinfo {volume} {49}},\ \bibinfo
  {pages} {2950} (\bibinfo {year} {1994})},\ \Eprint
  {http://arxiv.org/abs/nucl-th/9406039} {arXiv:nucl-th/9406039} \BibitemShut
  {NoStop}%
\bibitem [{\citenamefont {Wiringa}\ \emph {et~al.}(1995)\citenamefont
  {Wiringa}, \citenamefont {Stoks},\ and\ \citenamefont
  {Schiavilla}}]{Wiringa:1994wb}%
  \BibitemOpen
  \bibfield  {author} {\bibinfo {author} {\bibfnamefont {R.~B.}\ \bibnamefont
  {Wiringa}}, \bibinfo {author} {\bibfnamefont {V.}~\bibnamefont {Stoks}}, \
  and\ \bibinfo {author} {\bibfnamefont {R.}~\bibnamefont {Schiavilla}},\
  }\href {\doibase 10.1103/PhysRevC.51.38} {\bibfield  {journal} {\bibinfo
  {journal} {Phys. Rev. C}\ }\textbf {\bibinfo {volume} {51}},\ \bibinfo
  {pages} {38} (\bibinfo {year} {1995})},\ \Eprint
  {http://arxiv.org/abs/nucl-th/9408016} {arXiv:nucl-th/9408016} \BibitemShut
  {NoStop}%
\bibitem [{\citenamefont {Machleidt}(2001)}]{Machleidt:2000ge}%
  \BibitemOpen
  \bibfield  {author} {\bibinfo {author} {\bibfnamefont {R.}~\bibnamefont
  {Machleidt}},\ }\href {\doibase 10.1103/PhysRevC.63.024001} {\bibfield
  {journal} {\bibinfo  {journal} {Phys. Rev. C}\ }\textbf {\bibinfo {volume}
  {63}},\ \bibinfo {pages} {024001} (\bibinfo {year} {2001})},\ \Eprint
  {http://arxiv.org/abs/nucl-th/0006014} {arXiv:nucl-th/0006014} \BibitemShut
  {NoStop}%
\bibitem [{\citenamefont {Gross}\ and\ \citenamefont
  {Stadler}(2008)}]{Gross:2008ps}%
  \BibitemOpen
  \bibfield  {author} {\bibinfo {author} {\bibfnamefont {F.}~\bibnamefont
  {Gross}}\ and\ \bibinfo {author} {\bibfnamefont {A.}~\bibnamefont
  {Stadler}},\ }\href {\doibase 10.1103/PhysRevC.78.014005} {\bibfield
  {journal} {\bibinfo  {journal} {Phys. Rev. C}\ }\textbf {\bibinfo {volume}
  {78}},\ \bibinfo {pages} {014005} (\bibinfo {year} {2008})},\ \Eprint
  {http://arxiv.org/abs/0802.1552} {arXiv:0802.1552 [nucl-th]} \BibitemShut
  {NoStop}%
\bibitem [{\citenamefont {Navarro~Pérez}\ \emph
  {et~al.}(2013{\natexlab{a}})\citenamefont {Navarro~Pérez}, \citenamefont
  {Amaro},\ and\ \citenamefont {Ruiz~Arriola}}]{Perez:2013mwa}%
  \BibitemOpen
  \bibfield  {author} {\bibinfo {author} {\bibfnamefont {R.}~\bibnamefont
  {Navarro~Pérez}}, \bibinfo {author} {\bibfnamefont {J.}~\bibnamefont
  {Amaro}}, \ and\ \bibinfo {author} {\bibfnamefont {E.}~\bibnamefont
  {Ruiz~Arriola}},\ }\href {\doibase 10.1103/PhysRevC.88.024002} {\bibfield
  {journal} {\bibinfo  {journal} {Phys. Rev. C}\ }\textbf {\bibinfo {volume}
  {88}},\ \bibinfo {pages} {024002} (\bibinfo {year} {2013}{\natexlab{a}})},\
  \bibinfo {note} {[Erratum: Phys.Rev.C 88, 069902 (2013)]},\ \Eprint
  {http://arxiv.org/abs/1304.0895} {arXiv:1304.0895 [nucl-th]} \BibitemShut
  {NoStop}%
\bibitem [{\citenamefont {Navarro~Pérez}\ \emph
  {et~al.}(2014{\natexlab{a}})\citenamefont {Navarro~Pérez}, \citenamefont
  {Amaro},\ and\ \citenamefont {Ruiz~Arriola}}]{Perez:2013oba}%
  \BibitemOpen
  \bibfield  {author} {\bibinfo {author} {\bibfnamefont {R.}~\bibnamefont
  {Navarro~Pérez}}, \bibinfo {author} {\bibfnamefont {J.}~\bibnamefont
  {Amaro}}, \ and\ \bibinfo {author} {\bibfnamefont {E.}~\bibnamefont
  {Ruiz~Arriola}},\ }\href {\doibase 10.1103/PhysRevC.89.024004} {\bibfield
  {journal} {\bibinfo  {journal} {Phys. Rev. C}\ }\textbf {\bibinfo {volume}
  {89}},\ \bibinfo {pages} {024004} (\bibinfo {year} {2014}{\natexlab{a}})},\
  \Eprint {http://arxiv.org/abs/1310.6972} {arXiv:1310.6972 [nucl-th]}
  \BibitemShut {NoStop}%
\bibitem [{\citenamefont {Navarro~P\'erez}\ \emph
  {et~al.}(2014{\natexlab{a}})\citenamefont {Navarro~P\'erez}, \citenamefont
  {Amaro},\ and\ \citenamefont {Ruiz~Arriola}}]{Perez:2014yla}%
  \BibitemOpen
  \bibfield  {author} {\bibinfo {author} {\bibfnamefont {R.}~\bibnamefont
  {Navarro~P\'erez}}, \bibinfo {author} {\bibfnamefont {J.}~\bibnamefont
  {Amaro}}, \ and\ \bibinfo {author} {\bibfnamefont {E.}~\bibnamefont
  {Ruiz~Arriola}},\ }\href {\doibase 10.1103/PhysRevC.89.064006} {\bibfield
  {journal} {\bibinfo  {journal} {Phys. Rev. C}\ }\textbf {\bibinfo {volume}
  {89}},\ \bibinfo {pages} {064006} (\bibinfo {year} {2014}{\natexlab{a}})},\
  \Eprint {http://arxiv.org/abs/1404.0314} {arXiv:1404.0314 [nucl-th]}
  \BibitemShut {NoStop}%
\bibitem [{\citenamefont {Navarro~Pérez}\ \emph {et~al.}(2016)\citenamefont
  {Navarro~Pérez}, \citenamefont {Amaro},\ and\ \citenamefont
  {Ruiz~Arriola}}]{Perez:2014waa}%
  \BibitemOpen
  \bibfield  {author} {\bibinfo {author} {\bibfnamefont {R.}~\bibnamefont
  {Navarro~Pérez}}, \bibinfo {author} {\bibfnamefont {J.}~\bibnamefont
  {Amaro}}, \ and\ \bibinfo {author} {\bibfnamefont {E.}~\bibnamefont
  {Ruiz~Arriola}},\ }\href {\doibase 10.1088/0954-3899/43/11/114001} {\bibfield
   {journal} {\bibinfo  {journal} {J. Phys. G}\ }\textbf {\bibinfo {volume}
  {43}},\ \bibinfo {pages} {114001} (\bibinfo {year} {2016})},\ \Eprint
  {http://arxiv.org/abs/1410.8097} {arXiv:1410.8097 [nucl-th]} \BibitemShut
  {NoStop}%
\bibitem [{\citenamefont {Navarro~P\'erez}\ \emph {et~al.}(2017)\citenamefont
  {Navarro~P\'erez}, \citenamefont {Amaro},\ and\ \citenamefont
  {Ruiz~Arriola}}]{Perez:2016aol}%
  \BibitemOpen
  \bibfield  {author} {\bibinfo {author} {\bibfnamefont {R.}~\bibnamefont
  {Navarro~P\'erez}}, \bibinfo {author} {\bibfnamefont {J.}~\bibnamefont
  {Amaro}}, \ and\ \bibinfo {author} {\bibfnamefont {E.}~\bibnamefont
  {Ruiz~Arriola}},\ }\href {\doibase 10.1103/PhysRevC.95.064001} {\bibfield
  {journal} {\bibinfo  {journal} {Phys. Rev. C}\ }\textbf {\bibinfo {volume}
  {95}},\ \bibinfo {pages} {064001} (\bibinfo {year} {2017})},\ \Eprint
  {http://arxiv.org/abs/1606.00592} {arXiv:1606.00592 [nucl-th]} \BibitemShut
  {NoStop}%
\bibitem [{\citenamefont {Epelbaum}\ \emph {et~al.}(2020)\citenamefont
  {Epelbaum}, \citenamefont {Krebs},\ and\ \citenamefont
  {Reinert}}]{Epelbaum:2019kcf}%
  \BibitemOpen
  \bibfield  {author} {\bibinfo {author} {\bibfnamefont {E.}~\bibnamefont
  {Epelbaum}}, \bibinfo {author} {\bibfnamefont {H.}~\bibnamefont {Krebs}}, \
  and\ \bibinfo {author} {\bibfnamefont {P.}~\bibnamefont {Reinert}},\ }\href
  {\doibase 10.3389/fphy.2020.00098} {\bibfield  {journal} {\bibinfo  {journal}
  {Front. in Phys.}\ }\textbf {\bibinfo {volume} {8}},\ \bibinfo {pages} {98}
  (\bibinfo {year} {2020})},\ \Eprint {http://arxiv.org/abs/1911.11875}
  {arXiv:1911.11875 [nucl-th]} \BibitemShut {NoStop}%
\bibitem [{\citenamefont {Rodriguez~Entem}\ \emph {et~al.}(2020)\citenamefont
  {Rodriguez~Entem}, \citenamefont {Machleidt},\ and\ \citenamefont
  {Nosyk}}]{RodriguezEntem:2020jgp}%
  \BibitemOpen
  \bibfield  {author} {\bibinfo {author} {\bibfnamefont {D.}~\bibnamefont
  {Rodriguez~Entem}}, \bibinfo {author} {\bibfnamefont {R.}~\bibnamefont
  {Machleidt}}, \ and\ \bibinfo {author} {\bibfnamefont {Y.}~\bibnamefont
  {Nosyk}},\ }\href {\doibase 10.3389/fphy.2020.00057} {\bibfield  {journal}
  {\bibinfo  {journal} {Front. in Phys.}\ }\textbf {\bibinfo {volume} {8}},\
  \bibinfo {pages} {57} (\bibinfo {year} {2020})}\BibitemShut {NoStop}%
\bibitem [{\citenamefont {Machleidt}(2017)}]{Machleidt:2017vls}%
  \BibitemOpen
  \bibfield  {author} {\bibinfo {author} {\bibfnamefont {R.}~\bibnamefont
  {Machleidt}},\ }\href {\doibase 10.1142/S0218301317300053} {\bibfield
  {journal} {\bibinfo  {journal} {Int. J. Mod. Phys. E}\ }\textbf {\bibinfo
  {volume} {26}},\ \bibinfo {pages} {1730005} (\bibinfo {year} {2017})},\
  \Eprint {http://arxiv.org/abs/1710.07215} {arXiv:1710.07215 [nucl-th]}
  \BibitemShut {NoStop}%
\bibitem [{\citenamefont {Ruiz~Arriola}\ \emph {et~al.}(2020)\citenamefont
  {Ruiz~Arriola}, \citenamefont {Amaro},\ and\ \citenamefont
  {Navarro~Pérez}}]{RuizArriola:2019nnv}%
  \BibitemOpen
  \bibfield  {author} {\bibinfo {author} {\bibfnamefont {E.}~\bibnamefont
  {Ruiz~Arriola}}, \bibinfo {author} {\bibfnamefont {J.~E.}\ \bibnamefont
  {Amaro}}, \ and\ \bibinfo {author} {\bibfnamefont {R.}~\bibnamefont
  {Navarro~Pérez}},\ }\href {\doibase 10.3389/fphy.2020.00001} {\bibfield
  {journal} {\bibinfo  {journal} {Front. in Phys.}\ }\textbf {\bibinfo {volume}
  {8}},\ \bibinfo {pages} {1} (\bibinfo {year} {2020})},\ \Eprint
  {http://arxiv.org/abs/1911.09637} {arXiv:1911.09637 [nucl-th]} \BibitemShut
  {NoStop}%
\bibitem [{\citenamefont {Navarro~P\'erez}\ and\ \citenamefont
  {Ruiz~Arriola}(2020)}]{Perez:2020rtq}%
  \BibitemOpen
  \bibfield  {author} {\bibinfo {author} {\bibfnamefont {R.}~\bibnamefont
  {Navarro~P\'erez}}\ and\ \bibinfo {author} {\bibfnamefont {E.}~\bibnamefont
  {Ruiz~Arriola}},\ }\href {\doibase 10.1140/epja/s10050-020-00103-1}
  {\bibfield  {journal} {\bibinfo  {journal} {Eur. Phys. J. A}\ }\textbf
  {\bibinfo {volume} {56}},\ \bibinfo {pages} {99} (\bibinfo {year}
  {2020})}\BibitemShut {NoStop}%
\bibitem [{\citenamefont {Bryan}(1960)}]{bryan1960new}%
  \BibitemOpen
  \bibfield  {author} {\bibinfo {author} {\bibfnamefont {R.}~\bibnamefont
  {Bryan}},\ }\href@noop {} {\bibfield  {journal} {\bibinfo  {journal} {Il
  Nuovo Cimento (1955-1965)}\ }\textbf {\bibinfo {volume} {16}},\ \bibinfo
  {pages} {895} (\bibinfo {year} {1960})}\BibitemShut {NoStop}%
\bibitem [{\citenamefont {Hamada}\ and\ \citenamefont
  {Johnston}(1962)}]{Hamada:1962nq}%
  \BibitemOpen
  \bibfield  {author} {\bibinfo {author} {\bibfnamefont {T.}~\bibnamefont
  {Hamada}}\ and\ \bibinfo {author} {\bibfnamefont {I.}~\bibnamefont
  {Johnston}},\ }\href {\doibase 10.1016/0029-5582(62)90228-6} {\bibfield
  {journal} {\bibinfo  {journal} {Nucl. Phys.}\ }\textbf {\bibinfo {volume}
  {34}},\ \bibinfo {pages} {382} (\bibinfo {year} {1962})}\BibitemShut
  {NoStop}%
\bibitem [{\citenamefont {Reid}(1968)}]{Reid:1968sq}%
  \BibitemOpen
  \bibfield  {author} {\bibinfo {author} {\bibfnamefont {J.}~\bibnamefont
  {Reid}, \bibfnamefont {Roderick~V.}},\ }\href {\doibase
  10.1016/0003-4916(68)90126-7} {\bibfield  {journal} {\bibinfo  {journal}
  {Annals Phys.}\ }\textbf {\bibinfo {volume} {50}},\ \bibinfo {pages} {411}
  (\bibinfo {year} {1968})}\BibitemShut {NoStop}%
\bibitem [{\citenamefont {Signell}(1969)}]{signell1969nuclear}%
  \BibitemOpen
  \bibfield  {author} {\bibinfo {author} {\bibfnamefont {P.}~\bibnamefont
  {Signell}},\ }in\ \href@noop {} {\emph {\bibinfo {booktitle} {Advances in
  nuclear physics}}}\ (\bibinfo  {publisher} {Springer},\ \bibinfo {year}
  {1969})\ pp.\ \bibinfo {pages} {223--294}\BibitemShut {NoStop}%
\bibitem [{\citenamefont {Lacombe}\ \emph {et~al.}(1975)\citenamefont
  {Lacombe}, \citenamefont {Loiseau}, \citenamefont {Richard}, \citenamefont
  {Vinh~Mau}, \citenamefont {Pires},\ and\ \citenamefont
  {de~Tourreil}}]{Lacombe:1975qr}%
  \BibitemOpen
  \bibfield  {author} {\bibinfo {author} {\bibfnamefont {M.}~\bibnamefont
  {Lacombe}}, \bibinfo {author} {\bibfnamefont {B.}~\bibnamefont {Loiseau}},
  \bibinfo {author} {\bibfnamefont {J.-M.}\ \bibnamefont {Richard}}, \bibinfo
  {author} {\bibfnamefont {R.}~\bibnamefont {Vinh~Mau}}, \bibinfo {author}
  {\bibfnamefont {P.}~\bibnamefont {Pires}}, \ and\ \bibinfo {author}
  {\bibfnamefont {R.}~\bibnamefont {de~Tourreil}},\ }\href {\doibase
  10.1103/PhysRevD.12.1495} {\bibfield  {journal} {\bibinfo  {journal} {Phys.
  Rev. D}\ }\textbf {\bibinfo {volume} {12}},\ \bibinfo {pages} {1495}
  (\bibinfo {year} {1975})}\BibitemShut {NoStop}%
\bibitem [{\citenamefont {Lagaris}\ and\ \citenamefont
  {Pandharipande}(1981)}]{Lagaris:1981mm}%
  \BibitemOpen
  \bibfield  {author} {\bibinfo {author} {\bibfnamefont {I.}~\bibnamefont
  {Lagaris}}\ and\ \bibinfo {author} {\bibfnamefont {V.}~\bibnamefont
  {Pandharipande}},\ }\href {\doibase 10.1016/0375-9474(81)90240-2} {\bibfield
  {journal} {\bibinfo  {journal} {Nucl. Phys. A}\ }\textbf {\bibinfo {volume}
  {359}},\ \bibinfo {pages} {331} (\bibinfo {year} {1981})}\BibitemShut
  {NoStop}%
\bibitem [{\citenamefont {Wiringa}\ \emph {et~al.}(1984)\citenamefont
  {Wiringa}, \citenamefont {Smith},\ and\ \citenamefont
  {Ainsworth}}]{Wiringa:1984tg}%
  \BibitemOpen
  \bibfield  {author} {\bibinfo {author} {\bibfnamefont {R.~B.}\ \bibnamefont
  {Wiringa}}, \bibinfo {author} {\bibfnamefont {R.}~\bibnamefont {Smith}}, \
  and\ \bibinfo {author} {\bibfnamefont {T.}~\bibnamefont {Ainsworth}},\ }\href
  {\doibase 10.1103/PhysRevC.29.1207} {\bibfield  {journal} {\bibinfo
  {journal} {Phys. Rev. C}\ }\textbf {\bibinfo {volume} {29}},\ \bibinfo
  {pages} {1207} (\bibinfo {year} {1984})}\BibitemShut {NoStop}%
\bibitem [{\citenamefont {Ordonez}\ \emph {et~al.}(1996)\citenamefont
  {Ordonez}, \citenamefont {Ray},\ and\ \citenamefont {van
  Kolck}}]{Ordonez:1995rz}%
  \BibitemOpen
  \bibfield  {author} {\bibinfo {author} {\bibfnamefont {C.}~\bibnamefont
  {Ordonez}}, \bibinfo {author} {\bibfnamefont {L.}~\bibnamefont {Ray}}, \ and\
  \bibinfo {author} {\bibfnamefont {U.}~\bibnamefont {van Kolck}},\ }\href
  {\doibase 10.1103/PhysRevC.53.2086} {\bibfield  {journal} {\bibinfo
  {journal} {Phys. Rev. C}\ }\textbf {\bibinfo {volume} {53}},\ \bibinfo
  {pages} {2086} (\bibinfo {year} {1996})},\ \Eprint
  {http://arxiv.org/abs/hep-ph/9511380} {arXiv:hep-ph/9511380} \BibitemShut
  {NoStop}%
\bibitem [{\citenamefont {Navarro~P\'erez}\ \emph {et~al.}(2013)\citenamefont
  {Navarro~P\'erez}, \citenamefont {Amaro},\ and\ \citenamefont
  {Ruiz~Arriola}}]{NavarroPerez:2012qf}%
  \BibitemOpen
  \bibfield  {author} {\bibinfo {author} {\bibfnamefont {R.}~\bibnamefont
  {Navarro~P\'erez}}, \bibinfo {author} {\bibfnamefont {J.}~\bibnamefont
  {Amaro}}, \ and\ \bibinfo {author} {\bibfnamefont {E.}~\bibnamefont
  {Ruiz~Arriola}},\ }\href {\doibase 10.1016/j.physletb.2013.05.066} {\bibfield
   {journal} {\bibinfo  {journal} {Phys. Lett. B}\ }\textbf {\bibinfo {volume}
  {724}},\ \bibinfo {pages} {138} (\bibinfo {year} {2013})},\ \Eprint
  {http://arxiv.org/abs/1202.2689} {arXiv:1202.2689 [nucl-th]} \BibitemShut
  {NoStop}%
\bibitem [{\citenamefont {Epelbaum}\ \emph {et~al.}(2015)\citenamefont
  {Epelbaum}, \citenamefont {Krebs},\ and\ \citenamefont
  {Mei\ss{}ner}}]{Epelbaum:2014sza}%
  \BibitemOpen
  \bibfield  {author} {\bibinfo {author} {\bibfnamefont {E.}~\bibnamefont
  {Epelbaum}}, \bibinfo {author} {\bibfnamefont {H.}~\bibnamefont {Krebs}}, \
  and\ \bibinfo {author} {\bibfnamefont {U.}~\bibnamefont {Mei\ss{}ner}},\
  }\href {\doibase 10.1103/PhysRevLett.115.122301} {\bibfield  {journal}
  {\bibinfo  {journal} {Phys. Rev. Lett.}\ }\textbf {\bibinfo {volume} {115}},\
  \bibinfo {pages} {122301} (\bibinfo {year} {2015})},\ \Eprint
  {http://arxiv.org/abs/1412.4623} {arXiv:1412.4623 [nucl-th]} \BibitemShut
  {NoStop}%
\bibitem [{\citenamefont {Signell}\ and\ \citenamefont
  {Yoder}(1963)}]{signell1963comparison}%
  \BibitemOpen
  \bibfield  {author} {\bibinfo {author} {\bibfnamefont {P.}~\bibnamefont
  {Signell}}\ and\ \bibinfo {author} {\bibfnamefont {N.}~\bibnamefont
  {Yoder}},\ }\href@noop {} {\bibfield  {journal} {\bibinfo  {journal}
  {Physical Review}\ }\textbf {\bibinfo {volume} {132}},\ \bibinfo {pages}
  {1707} (\bibinfo {year} {1963})}\BibitemShut {NoStop}%
\bibitem [{\citenamefont {Signell}\ and\ \citenamefont
  {Yoder}(1964)}]{signell1964comparison}%
  \BibitemOpen
  \bibfield  {author} {\bibinfo {author} {\bibfnamefont {P.}~\bibnamefont
  {Signell}}\ and\ \bibinfo {author} {\bibfnamefont {N.}~\bibnamefont
  {Yoder}},\ }\href@noop {} {\bibfield  {journal} {\bibinfo  {journal}
  {Physical Review}\ }\textbf {\bibinfo {volume} {134}},\ \bibinfo {pages}
  {B100} (\bibinfo {year} {1964})}\BibitemShut {NoStop}%
\bibitem [{\citenamefont {Nagels}\ \emph {et~al.}(1975)\citenamefont {Nagels},
  \citenamefont {Rijken},\ and\ \citenamefont {de~Swart}}]{Nagels:1975fb}%
  \BibitemOpen
  \bibfield  {author} {\bibinfo {author} {\bibfnamefont {M.}~\bibnamefont
  {Nagels}}, \bibinfo {author} {\bibfnamefont {T.}~\bibnamefont {Rijken}}, \
  and\ \bibinfo {author} {\bibfnamefont {J.}~\bibnamefont {de~Swart}},\ }\href
  {\doibase 10.1103/PhysRevD.12.744} {\bibfield  {journal} {\bibinfo  {journal}
  {Phys. Rev. D}\ }\textbf {\bibinfo {volume} {12}},\ \bibinfo {pages} {744}
  (\bibinfo {year} {1975})}\BibitemShut {NoStop}%
\bibitem [{\citenamefont {Nagels}\ \emph {et~al.}(1978)\citenamefont {Nagels},
  \citenamefont {Rijken},\ and\ \citenamefont {de~Swart}}]{Nagels:1977ze}%
  \BibitemOpen
  \bibfield  {author} {\bibinfo {author} {\bibfnamefont {M.}~\bibnamefont
  {Nagels}}, \bibinfo {author} {\bibfnamefont {T.}~\bibnamefont {Rijken}}, \
  and\ \bibinfo {author} {\bibfnamefont {J.}~\bibnamefont {de~Swart}},\ }\href
  {\doibase 10.1103/PhysRevD.17.768} {\bibfield  {journal} {\bibinfo  {journal}
  {Phys. Rev. D}\ }\textbf {\bibinfo {volume} {17}},\ \bibinfo {pages} {768}
  (\bibinfo {year} {1978})}\BibitemShut {NoStop}%
\bibitem [{\citenamefont {Nagels}\ \emph {et~al.}(1979)\citenamefont {Nagels},
  \citenamefont {Rijken},\ and\ \citenamefont {de~Swart}}]{Nagels:1978sc}%
  \BibitemOpen
  \bibfield  {author} {\bibinfo {author} {\bibfnamefont {M.}~\bibnamefont
  {Nagels}}, \bibinfo {author} {\bibfnamefont {T.}~\bibnamefont {Rijken}}, \
  and\ \bibinfo {author} {\bibfnamefont {J.}~\bibnamefont {de~Swart}},\ }\href
  {\doibase 10.1103/PhysRevD.20.1633} {\bibfield  {journal} {\bibinfo
  {journal} {Phys. Rev. D}\ }\textbf {\bibinfo {volume} {20}},\ \bibinfo
  {pages} {1633} (\bibinfo {year} {1979})}\BibitemShut {NoStop}%
\bibitem [{\citenamefont {Navarro~P\'erez}\ \emph
  {et~al.}(2014{\natexlab{b}})\citenamefont {Navarro~P\'erez}, \citenamefont
  {Amaro},\ and\ \citenamefont {Ruiz~Arriola}}]{NavarroPerez:2014ihw}%
  \BibitemOpen
  \bibfield  {author} {\bibinfo {author} {\bibfnamefont {R.}~\bibnamefont
  {Navarro~P\'erez}}, \bibinfo {author} {\bibfnamefont {J.~E.}\ \bibnamefont
  {Amaro}}, \ and\ \bibinfo {author} {\bibfnamefont {E.}~\bibnamefont
  {Ruiz~Arriola}},\ }\href {\doibase 10.1103/PhysRevC.89.064006} {\bibfield
  {journal} {\bibinfo  {journal} {Phys. Rev. C}\ }\textbf {\bibinfo {volume}
  {89}},\ \bibinfo {pages} {064006} (\bibinfo {year} {2014}{\natexlab{b}})},\
  \Eprint {http://arxiv.org/abs/1404.0314} {arXiv:1404.0314 [nucl-th]}
  \BibitemShut {NoStop}%
\bibitem [{\citenamefont {Navarro~P\'erez}\ \emph {et~al.}(2015)\citenamefont
  {Navarro~P\'erez}, \citenamefont {Amaro},\ and\ \citenamefont
  {Ruiz~Arriola}}]{NavarroPerez:2014rvx}%
  \BibitemOpen
  \bibfield  {author} {\bibinfo {author} {\bibfnamefont {R.}~\bibnamefont
  {Navarro~P\'erez}}, \bibinfo {author} {\bibfnamefont {J.~E.}\ \bibnamefont
  {Amaro}}, \ and\ \bibinfo {author} {\bibfnamefont {E.}~\bibnamefont
  {Ruiz~Arriola}},\ }\href {\doibase 10.1088/0954-3899/42/3/034013} {\bibfield
  {journal} {\bibinfo  {journal} {J. Phys. G}\ }\textbf {\bibinfo {volume}
  {42}},\ \bibinfo {pages} {034013} (\bibinfo {year} {2015})},\ \Eprint
  {http://arxiv.org/abs/1406.0625} {arXiv:1406.0625 [nucl-th]} \BibitemShut
  {NoStop}%
\bibitem [{\citenamefont {Navarro~P\'erez}\ \emph
  {et~al.}(2016{\natexlab{a}})\citenamefont {Navarro~P\'erez}, \citenamefont
  {Amaro},\ and\ \citenamefont {Ruiz~Arriola}}]{NavarroPerez:2016wqg}%
  \BibitemOpen
  \bibfield  {author} {\bibinfo {author} {\bibfnamefont {R.}~\bibnamefont
  {Navarro~P\'erez}}, \bibinfo {author} {\bibfnamefont {J.~E.}\ \bibnamefont
  {Amaro}}, \ and\ \bibinfo {author} {\bibfnamefont {E.}~\bibnamefont
  {Ruiz~Arriola}},\ }\href {\doibase 10.1142/S0218301316410093} {\bibfield
  {journal} {\bibinfo  {journal} {Int. J. Mod. Phys. E}\ }\textbf {\bibinfo
  {volume} {25}},\ \bibinfo {pages} {1641009} (\bibinfo {year}
  {2016}{\natexlab{a}})},\ \Eprint {http://arxiv.org/abs/1601.08220}
  {arXiv:1601.08220 [nucl-th]} \BibitemShut {NoStop}%
\bibitem [{\citenamefont {Simo}\ \emph {et~al.}(2018)\citenamefont {Simo},
  \citenamefont {Amaro}, \citenamefont {Ruiz~Arriola},\ and\ \citenamefont
  {Navarro~Pérez}}]{RuizSimo:2017anp}%
  \BibitemOpen
  \bibfield  {author} {\bibinfo {author} {\bibfnamefont {I.}~\bibnamefont
  {Simo}}, \bibinfo {author} {\bibfnamefont {J.}~\bibnamefont {Amaro}},
  \bibinfo {author} {\bibfnamefont {E.}~\bibnamefont {Ruiz~Arriola}}, \ and\
  \bibinfo {author} {\bibfnamefont {R.}~\bibnamefont {Navarro~Pérez}},\ }\href
  {\doibase 10.1088/1361-6471/aaabd2} {\bibfield  {journal} {\bibinfo
  {journal} {J. Phys. G}\ }\textbf {\bibinfo {volume} {45}},\ \bibinfo {pages}
  {035107} (\bibinfo {year} {2018})},\ \Eprint
  {http://arxiv.org/abs/1705.06522} {arXiv:1705.06522 [nucl-th]} \BibitemShut
  {NoStop}%
\bibitem [{\citenamefont {Carlsson}\ \emph {et~al.}(2016)\citenamefont
  {Carlsson}, \citenamefont {Ekstr\"om}, \citenamefont {Forss\'en},
  \citenamefont {Str\"omberg}, \citenamefont {Jansen}, \citenamefont {Lilja},
  \citenamefont {Lindby}, \citenamefont {Mattsson},\ and\ \citenamefont
  {Wendt}}]{Carlsson:2015vda}%
  \BibitemOpen
  \bibfield  {author} {\bibinfo {author} {\bibfnamefont {B.}~\bibnamefont
  {Carlsson}}, \bibinfo {author} {\bibfnamefont {A.}~\bibnamefont {Ekstr\"om}},
  \bibinfo {author} {\bibfnamefont {C.}~\bibnamefont {Forss\'en}}, \bibinfo
  {author} {\bibfnamefont {D.}~\bibnamefont {Str\"omberg}}, \bibinfo {author}
  {\bibfnamefont {G.}~\bibnamefont {Jansen}}, \bibinfo {author} {\bibfnamefont
  {O.}~\bibnamefont {Lilja}}, \bibinfo {author} {\bibfnamefont
  {M.}~\bibnamefont {Lindby}}, \bibinfo {author} {\bibfnamefont
  {B.}~\bibnamefont {Mattsson}}, \ and\ \bibinfo {author} {\bibfnamefont
  {K.}~\bibnamefont {Wendt}},\ }\href {\doibase 10.1103/PhysRevX.6.011019}
  {\bibfield  {journal} {\bibinfo  {journal} {Phys. Rev. X}\ }\textbf {\bibinfo
  {volume} {6}},\ \bibinfo {pages} {011019} (\bibinfo {year} {2016})},\ \Eprint
  {http://arxiv.org/abs/1506.02466} {arXiv:1506.02466 [nucl-th]} \BibitemShut
  {NoStop}%
\bibitem [{\citenamefont {Beane}\ \emph {et~al.}(2006)\citenamefont {Beane},
  \citenamefont {Bedaque}, \citenamefont {Orginos},\ and\ \citenamefont
  {Savage}}]{Beane:2006mx}%
  \BibitemOpen
  \bibfield  {author} {\bibinfo {author} {\bibfnamefont {S.}~\bibnamefont
  {Beane}}, \bibinfo {author} {\bibfnamefont {P.}~\bibnamefont {Bedaque}},
  \bibinfo {author} {\bibfnamefont {K.}~\bibnamefont {Orginos}}, \ and\
  \bibinfo {author} {\bibfnamefont {M.}~\bibnamefont {Savage}},\ }\href
  {\doibase 10.1103/PhysRevLett.97.012001} {\bibfield  {journal} {\bibinfo
  {journal} {Phys. Rev. Lett.}\ }\textbf {\bibinfo {volume} {97}},\ \bibinfo
  {pages} {012001} (\bibinfo {year} {2006})},\ \Eprint
  {http://arxiv.org/abs/hep-lat/0602010} {arXiv:hep-lat/0602010} \BibitemShut
  {NoStop}%
\bibitem [{\citenamefont {Aoki}(2013)}]{Aoki:2013tba}%
  \BibitemOpen
  \bibfield  {author} {\bibinfo {author} {\bibfnamefont {S.}~\bibnamefont
  {Aoki}},\ }\href {\doibase 10.1140/epja/i2013-13081-0} {\bibfield  {journal}
  {\bibinfo  {journal} {Eur. Phys. J. A}\ }\textbf {\bibinfo {volume} {49}},\
  \bibinfo {pages} {81} (\bibinfo {year} {2013})},\ \Eprint
  {http://arxiv.org/abs/1309.4150} {arXiv:1309.4150 [hep-lat]} \BibitemShut
  {NoStop}%
\bibitem [{\citenamefont {H\"orz}\ \emph {et~al.}(2020)\citenamefont {H\"orz}
  \emph {et~al.}}]{Horz:2020zvv}%
  \BibitemOpen
  \bibfield  {author} {\bibinfo {author} {\bibfnamefont {B.}~\bibnamefont
  {H\"orz}} \emph {et~al.},\ }\href@noop {} {\  (\bibinfo {year} {2020})},\
  \Eprint {http://arxiv.org/abs/2009.11825} {arXiv:2009.11825 [hep-lat]}
  \BibitemShut {NoStop}%
\bibitem [{\citenamefont {Illa}\ \emph {et~al.}(2020)\citenamefont {Illa} \emph
  {et~al.}}]{Illa:2020nsi}%
  \BibitemOpen
  \bibfield  {author} {\bibinfo {author} {\bibfnamefont {M.}~\bibnamefont
  {Illa}} \emph {et~al.},\ }\href@noop {} {\  (\bibinfo {year} {2020})},\
  \Eprint {http://arxiv.org/abs/2009.12357} {arXiv:2009.12357 [hep-lat]}
  \BibitemShut {NoStop}%
\bibitem [{\citenamefont {Drischler}\ \emph {et~al.}(2019)\citenamefont
  {Drischler}, \citenamefont {Haxton}, \citenamefont {McElvain}, \citenamefont
  {Mereghetti}, \citenamefont {Nicholson}, \citenamefont {Vranas},\ and\
  \citenamefont {Walker-Loud}}]{Drischler:2019xuo}%
  \BibitemOpen
  \bibfield  {author} {\bibinfo {author} {\bibfnamefont {C.}~\bibnamefont
  {Drischler}}, \bibinfo {author} {\bibfnamefont {W.}~\bibnamefont {Haxton}},
  \bibinfo {author} {\bibfnamefont {K.}~\bibnamefont {McElvain}}, \bibinfo
  {author} {\bibfnamefont {E.}~\bibnamefont {Mereghetti}}, \bibinfo {author}
  {\bibfnamefont {A.}~\bibnamefont {Nicholson}}, \bibinfo {author}
  {\bibfnamefont {P.}~\bibnamefont {Vranas}}, \ and\ \bibinfo {author}
  {\bibfnamefont {A.}~\bibnamefont {Walker-Loud}}\ }(\bibinfo {year} {2019})\
  \Eprint {http://arxiv.org/abs/1910.07961} {arXiv:1910.07961 [nucl-th]}
  \BibitemShut {NoStop}%
\bibitem [{\citenamefont {Arndt}\ and\ \citenamefont
  {Mac~Gregor}(1966)}]{Arndt:1966in}%
  \BibitemOpen
  \bibfield  {author} {\bibinfo {author} {\bibfnamefont {R.~A.}\ \bibnamefont
  {Arndt}}\ and\ \bibinfo {author} {\bibfnamefont {M.~H.}\ \bibnamefont
  {Mac~Gregor}},\ }\href {\doibase 10.1103/PhysRev.141.873} {\bibfield
  {journal} {\bibinfo  {journal} {Phys. Rev.}\ }\textbf {\bibinfo {volume}
  {141}},\ \bibinfo {pages} {873} (\bibinfo {year} {1966})}\BibitemShut
  {NoStop}%
\bibitem [{\citenamefont {Bystricky}\ \emph {et~al.}(1978)\citenamefont
  {Bystricky}, \citenamefont {Lehar},\ and\ \citenamefont
  {Winternitz}}]{Bystricky:1976jr}%
  \BibitemOpen
  \bibfield  {author} {\bibinfo {author} {\bibfnamefont {J.}~\bibnamefont
  {Bystricky}}, \bibinfo {author} {\bibfnamefont {F.}~\bibnamefont {Lehar}}, \
  and\ \bibinfo {author} {\bibfnamefont {P.}~\bibnamefont {Winternitz}},\
  }\href {\doibase 10.1051/jphys:019780039010100} {\bibfield  {journal}
  {\bibinfo  {journal} {J. Phys. (France)}\ }\textbf {\bibinfo {volume} {39}},\
  \bibinfo {pages} {1} (\bibinfo {year} {1978})}\BibitemShut {NoStop}%
\bibitem [{\citenamefont {La~France}\ and\ \citenamefont
  {Winternitz}(1980)}]{la1980scattering}%
  \BibitemOpen
  \bibfield  {author} {\bibinfo {author} {\bibfnamefont {P.}~\bibnamefont
  {La~France}}\ and\ \bibinfo {author} {\bibfnamefont {P.}~\bibnamefont
  {Winternitz}},\ }\href@noop {} {\bibfield  {journal} {\bibinfo  {journal}
  {Journal de Physique}\ }\textbf {\bibinfo {volume} {41}},\ \bibinfo {pages}
  {1391} (\bibinfo {year} {1980})}\BibitemShut {NoStop}%
\bibitem [{\citenamefont {Kamada}\ \emph {et~al.}(2011)\citenamefont {Kamada},
  \citenamefont {Gl{\"o}ckle}, \citenamefont {Wita{\l}a}, \citenamefont
  {Golak},\ and\ \citenamefont {Skibi{\'n}ski}}]{kamada2011determination}%
  \BibitemOpen
  \bibfield  {author} {\bibinfo {author} {\bibfnamefont {H.}~\bibnamefont
  {Kamada}}, \bibinfo {author} {\bibfnamefont {W.}~\bibnamefont {Gl{\"o}ckle}},
  \bibinfo {author} {\bibfnamefont {H.}~\bibnamefont {Wita{\l}a}}, \bibinfo
  {author} {\bibfnamefont {J.}~\bibnamefont {Golak}}, \ and\ \bibinfo {author}
  {\bibfnamefont {R.}~\bibnamefont {Skibi{\'n}ski}},\ }\href@noop {} {\bibfield
   {journal} {\bibinfo  {journal} {Few-Body Systems}\ }\textbf {\bibinfo
  {volume} {50}},\ \bibinfo {pages} {231} (\bibinfo {year} {2011})}\BibitemShut
  {NoStop}%
\bibitem [{\citenamefont {Navarro~Pérez}\ \emph
  {et~al.}(2013{\natexlab{b}})\citenamefont {Navarro~Pérez}, \citenamefont
  {Amaro},\ and\ \citenamefont {Ruiz~Arriola}}]{Perez:2013jpa}%
  \BibitemOpen
  \bibfield  {author} {\bibinfo {author} {\bibfnamefont {R.}~\bibnamefont
  {Navarro~Pérez}}, \bibinfo {author} {\bibfnamefont {J.}~\bibnamefont
  {Amaro}}, \ and\ \bibinfo {author} {\bibfnamefont {E.}~\bibnamefont
  {Ruiz~Arriola}},\ }\href {\doibase 10.1103/PhysRevC.88.064002} {\bibfield
  {journal} {\bibinfo  {journal} {Phys. Rev. C}\ }\textbf {\bibinfo {volume}
  {88}},\ \bibinfo {pages} {064002} (\bibinfo {year} {2013}{\natexlab{b}})},\
  \bibinfo {note} {[Erratum: Phys.Rev.C 91, 029901 (2015)]},\ \Eprint
  {http://arxiv.org/abs/1310.2536} {arXiv:1310.2536 [nucl-th]} \BibitemShut
  {NoStop}%
\bibitem [{\citenamefont {Ruiz~de Elvira}\ and\ \citenamefont
  {Ruiz~Arriola}(2018)}]{RuizdeElvira:2018hsv}%
  \BibitemOpen
  \bibfield  {author} {\bibinfo {author} {\bibfnamefont {J.}~\bibnamefont
  {Ruiz~de Elvira}}\ and\ \bibinfo {author} {\bibfnamefont {E.}~\bibnamefont
  {Ruiz~Arriola}},\ }\href {\doibase 10.1140/epjc/s10052-018-6342-7} {\bibfield
   {journal} {\bibinfo  {journal} {Eur. Phys. J. C}\ }\textbf {\bibinfo
  {volume} {78}},\ \bibinfo {pages} {878} (\bibinfo {year} {2018})},\ \Eprint
  {http://arxiv.org/abs/1807.10837} {arXiv:1807.10837 [hep-ph]} \BibitemShut
  {NoStop}%
\bibitem [{\citenamefont {Nagels}\ \emph {et~al.}(2019)\citenamefont {Nagels},
  \citenamefont {Rijken},\ and\ \citenamefont {Yamamoto}}]{Nagels:2014qqa}%
  \BibitemOpen
  \bibfield  {author} {\bibinfo {author} {\bibfnamefont {M.~M.}\ \bibnamefont
  {Nagels}}, \bibinfo {author} {\bibfnamefont {T.~A.}\ \bibnamefont {Rijken}},
  \ and\ \bibinfo {author} {\bibfnamefont {Y.}~\bibnamefont {Yamamoto}},\
  }\href {\doibase 10.1103/PhysRevC.99.044002} {\bibfield  {journal} {\bibinfo
  {journal} {Phys. Rev. C}\ }\textbf {\bibinfo {volume} {99}},\ \bibinfo
  {pages} {044002} (\bibinfo {year} {2019})},\ \Eprint
  {http://arxiv.org/abs/1408.4825} {arXiv:1408.4825 [nucl-th]} \BibitemShut
  {NoStop}%
\bibitem [{\citenamefont {Navarro~P\'erez}\ \emph
  {et~al.}(2016{\natexlab{b}})\citenamefont {Navarro~P\'erez}, \citenamefont
  {Amaro},\ and\ \citenamefont {Ruiz~Arriola}}]{Perez:2016vzj}%
  \BibitemOpen
  \bibfield  {author} {\bibinfo {author} {\bibfnamefont {R.}~\bibnamefont
  {Navarro~P\'erez}}, \bibinfo {author} {\bibfnamefont {J.}~\bibnamefont
  {Amaro}}, \ and\ \bibinfo {author} {\bibfnamefont {E.}~\bibnamefont
  {Ruiz~Arriola}},\ }\href {\doibase 10.1142/S0218301316410093} {\bibfield
  {journal} {\bibinfo  {journal} {Int. J. Mod. Phys. E}\ }\textbf {\bibinfo
  {volume} {25}},\ \bibinfo {pages} {1641009} (\bibinfo {year}
  {2016}{\natexlab{b}})},\ \Eprint {http://arxiv.org/abs/1601.08220}
  {arXiv:1601.08220 [nucl-th]} \BibitemShut {NoStop}%
\bibitem [{\citenamefont {Taylor}(1997)}]{taylor1997introduction}%
  \BibitemOpen
  \bibfield  {author} {\bibinfo {author} {\bibfnamefont {J.}~\bibnamefont
  {Taylor}},\ }\href@noop {} {\emph {\bibinfo {title} {Introduction to error
  analysis, the study of uncertainties in physical measurements}}}\ (\bibinfo
  {year} {1997})\BibitemShut {NoStop}%
\bibitem [{\citenamefont {Evans}\ and\ \citenamefont
  {Rosenthal}(2004)}]{evans2004probability}%
  \BibitemOpen
  \bibfield  {author} {\bibinfo {author} {\bibfnamefont {M.~J.}\ \bibnamefont
  {Evans}}\ and\ \bibinfo {author} {\bibfnamefont {J.~S.}\ \bibnamefont
  {Rosenthal}},\ }\href@noop {} {\emph {\bibinfo {title} {Probability and
  statistics: The science of uncertainty}}}\ (\bibinfo  {publisher}
  {Macmillan},\ \bibinfo {year} {2004})\BibitemShut {NoStop}%
\bibitem [{\citenamefont {Eadie}\ and\ \citenamefont
  {James}(2006)}]{eadie2006statistical}%
  \BibitemOpen
  \bibfield  {author} {\bibinfo {author} {\bibfnamefont {W.~T.}\ \bibnamefont
  {Eadie}}\ and\ \bibinfo {author} {\bibfnamefont {F.}~\bibnamefont {James}},\
  }\href@noop {} {\emph {\bibinfo {title} {Statistical methods in experimental
  physics}}}\ (\bibinfo  {publisher} {World Scientific},\ \bibinfo {year}
  {2006})\BibitemShut {NoStop}%
\bibitem [{\citenamefont {Vrbik}(2005)}]{vrbik2005population}%
  \BibitemOpen
  \bibfield  {author} {\bibinfo {author} {\bibfnamefont {J.}~\bibnamefont
  {Vrbik}},\ }\href@noop {} {\bibfield  {journal} {\bibinfo  {journal}
  {Computational Statistics}\ }\textbf {\bibinfo {volume} {20}},\ \bibinfo
  {pages} {611} (\bibinfo {year} {2005})}\BibitemShut {NoStop}%
\bibitem [{\citenamefont {Navarro~Pérez}\ \emph
  {et~al.}(2014{\natexlab{b}})\citenamefont {Navarro~Pérez}, \citenamefont
  {Amaro},\ and\ \citenamefont {Ruiz~Arriola}}]{Perez:2014jsa}%
  \BibitemOpen
  \bibfield  {author} {\bibinfo {author} {\bibfnamefont {R.}~\bibnamefont
  {Navarro~Pérez}}, \bibinfo {author} {\bibfnamefont {J.}~\bibnamefont
  {Amaro}}, \ and\ \bibinfo {author} {\bibfnamefont {E.}~\bibnamefont
  {Ruiz~Arriola}},\ }\href {\doibase 10.1016/j.physletb.2014.09.035} {\bibfield
   {journal} {\bibinfo  {journal} {Phys. Lett. B}\ }\textbf {\bibinfo {volume}
  {738}},\ \bibinfo {pages} {155} (\bibinfo {year} {2014}{\natexlab{b}})},\
  \Eprint {http://arxiv.org/abs/1407.3937} {arXiv:1407.3937 [nucl-th]}
  \BibitemShut {NoStop}%
\bibitem [{\citenamefont {Hotelling}(1953)}]{hotelling1953new}%
  \BibitemOpen
  \bibfield  {author} {\bibinfo {author} {\bibfnamefont {H.}~\bibnamefont
  {Hotelling}},\ }\href@noop {} {\bibfield  {journal} {\bibinfo  {journal}
  {Journal of the Royal Statistical Society. Series B (Methodological)}\
  }\textbf {\bibinfo {volume} {15}},\ \bibinfo {pages} {193} (\bibinfo {year}
  {1953})}\BibitemShut {NoStop}%
\end{thebibliography}%

\end{document}